\documentclass[aps,prl,reprint,superscriptaddress,showpacs]{revtex4-1}
\usepackage{graphicx}% Include figure files
\usepackage{amssymb}
\usepackage{amsmath}
\usepackage{bm}
\usepackage{verbatim}
\usepackage[bookmarks=false,colorlinks=true,urlcolor=blue,citecolor=blue,linkcolor=blue]{hyperref}
\usepackage[normalem]{ulem}

\renewcommand{\vec}[1]{{\bm #1}}

\renewcommand{\bf}[1]{{\textbf #1}}

\begin{document}

\title{Cooper Pair Induced Frustration and Nematicity of Two-Dimensional Magnetic Adatom Lattices}

\author{Michael Schecter}
\affiliation{Center for Quantum Devices, Niels Bohr Institute, University of Copenhagen, 2100 Copenhagen, Denmark}
\affiliation{Condensed Matter Theory Center and Joint Quantum Institute, Department of Physics, University of Maryland, College Park, Maryland 20742, USA}

\author{Olav Sylju{\aa}sen}
\affiliation{Department of Physics, University of Oslo, P. O. Box 1048 Blindern, N-0316 Oslo, Norway}

\author{Jens Paaske}
\affiliation{Center for Quantum Devices, Niels Bohr Institute, University of Copenhagen, 2100 Copenhagen, Denmark}

\date{\today}

\begin{abstract}
We propose utilizing the Cooper pair to induce magnetic frustration in systems of two-dimensional (2D) magnetic adatom lattices on $s$-wave superconducting surfaces.
The competition between singlet electron correlations and the RKKY coupling is shown to lead to a variety of hidden order states that break the point-group symmetry of the 2D adatom lattice at finite temperature. The phase diagram is constructed using a newly developed effective bond theory [M. Schecter et al., Phys. Rev. Lett. \textbf{119}, 157202 (2017)], and exhibits broad regions of long-range vestigial nematic order.

\end{abstract}


\maketitle

The interplay between magnetism and superconductivity has a long and rich history, sometimes yielding novel forms of matter with intertwined or competing orders. A striking example may occur for systems of magnetic adatoms exchange coupled to a superconducting surface \cite{Fu07,Ji08,Franke11,Fu12,Nadj-Perge14,Hatter15,Ruby15,Ruby16,Pawlak16,Heinrich17,Kezilebieke17,Ruby17,Choi17,Hatter17}, which could provide a route towards creating interfacial topological phases harboring Majorana bound states \cite{Nadj-Perge14,Ruby15,Pawlak16,Pientka13,Pientka14,Peng15,Klinovaja13,Braunecker13,
Vazifeh13,Kim14,Heimes14,Heimes15,Reis14,Brydon15,Li14,Schecter15,Choy11,
Martin12}.

\begin{figure}[t]
\includegraphics[width=\columnwidth]{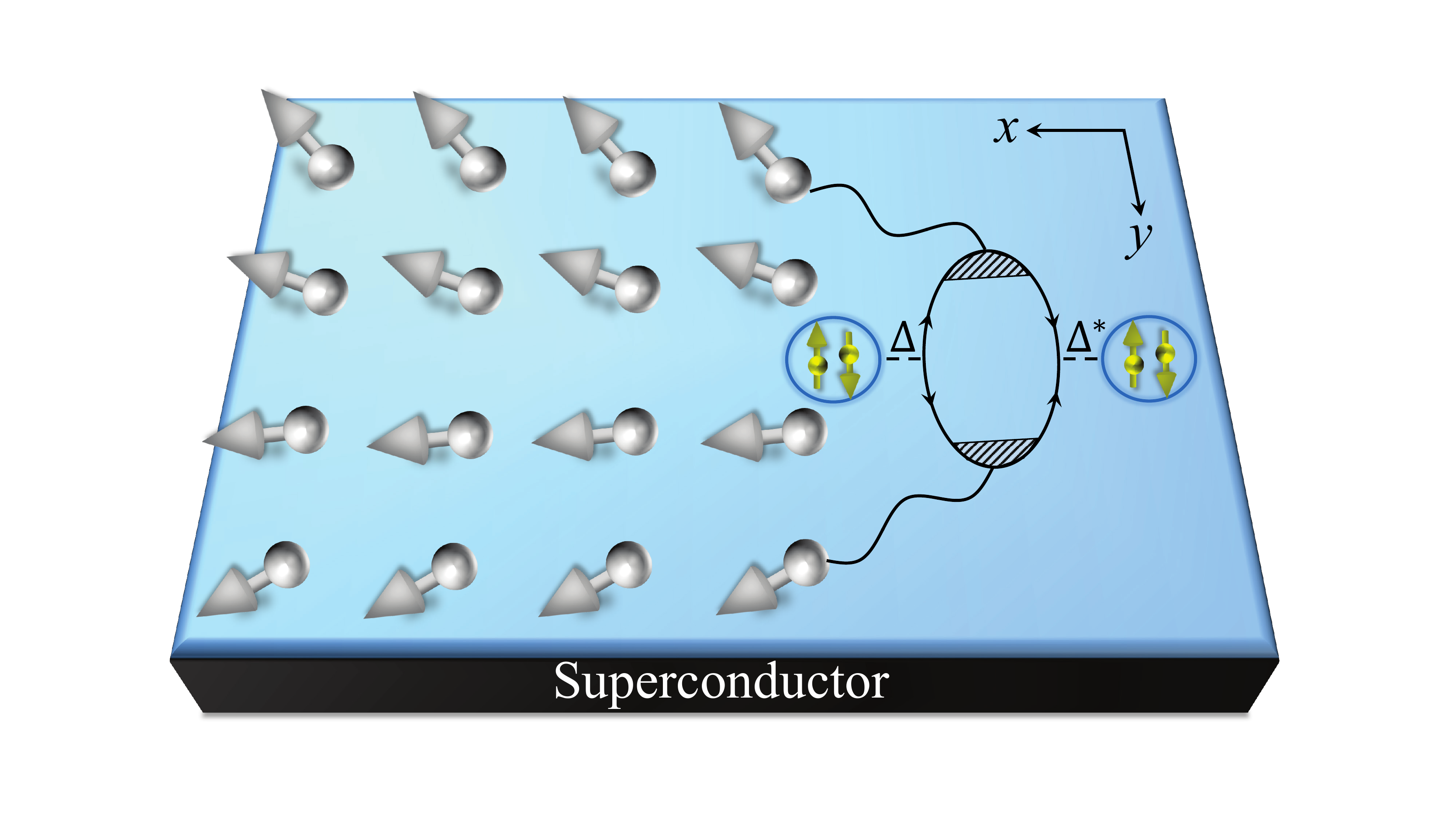}
\caption{A 2D magnetic adatom lattice is exchange coupled to a superconducting surface. Cooper pairs in the bulk mediate an indirect exchange coupling between adatoms that can frustrate the RKKY coupling (not depicted). The Cooper pair contribution is enhanced by YSR states induced by the adatoms (shaded vertex). This competition can drive spontaneous point group symmetry breaking of the adatom lattice at finite temperatures, giving rise to states with long range lattice-nematic order.}
\label{fig:1}
\end{figure}

While some theoretical studies have considered the topological superconducting phase diagram for a few hand-selected 2D magnetic configurations \cite{Nakosai13,Rontynen15,Rontynen16}, little is known about the actual low temperature magnetic phase diagram. This problem is nontrivial due to the magnetic exchange frustration created by the competition between the Ruderman-Kittel-Kosuya-Yosida \cite{Ruderman54,Kasuya56,Yosida57} (RKKY) coupling and the antiferromagnetic (AFM) coupling mediated by singlet Cooper pairs \cite{Abrikosov88,Aristov97,Galitski02,Yao14,Schecter16}, see Fig.~\ref{fig:1}. While the magnitude of the Cooper pair contribution is small, scaling with the superconducting gap $\Delta$, it is longer ranged than the RKKY component and can lead to an instability of a FM chain towards helimagnetism \cite{Schecter16,Christensen16}.

For 2D adatom lattices, the helimagnetic wavevector ${\bf Q}$ is accompanied by a discrete set of symmetry-related wavevectors in the ground state manifold. This discrete symmetry is expressed through the exchange coupling bonds of the lattice and can be broken spontaneously by the spins at a {\it finite} temperature \cite{Chandra90,Capriotti04,Weber03}. The remarkable possibility of breaking a discrete symmetry with degrees of freedom that have only a \emph{continuous} local symmetry is one of the hallmark predictions of the ``order by disorder" mechanism \cite{Villain77,Henley89,Chandra90}. Here we investigate this phenomenon using a newly developed effective exchange bond theory \cite{Schecter17}, which generically predicts short-range helimagnetic states with long-range vestigial lattice-nematic order.

%%%%%%%%%%%
%%%%%%%%%%%
%%%%%%%%%%%
%%%%%%%%%%%

In particular, we construct the phase diagram of a square adatom lattice exchange coupled to a 3D superconductor in the plane of temperature $T$ and adatom lattice constant $a$, using the effective exchange bond theory of Ref.~\cite{Schecter17}. We find broad regions of symmetry broken phases in the regime where the RKKY coupling is FM (tuned by the lattice constant $a$). The spin-spin correlation length $\xi$, while always finite for $T>0$, exhibits a strong nonanalytic increase as the system enters the symmetry broken phase and is accompanied by anisotropic spin-spin correlations. Our results suggest that magnetic adatoms on superconducting surfaces provide a novel setting for the study of frustrated magnetism.

We describe the system of magnetic adatoms coupled to a 3D superconductor using the Bogoliubov-de Gennes Hamiltonian

\begin{equation}
\label{eq:H}
H=\frac{1}{2}\sum_\mathbf{k}\Psi_\mathbf{k}^\dagger\left(\xi_\mathbf{k}\tau_z+\Delta\tau_x\right)\Psi_\mathbf{k}
+\frac{1}{2}J\int_\mathbf{r}\Psi_\mathbf{r}^\dagger \vec{S}_\mathbf{r}\cdot\boldsymbol{\sigma}\Psi_\mathbf{r},
\end{equation}
where $\xi(\mathbf{k})=\frac{\mathbf{k}^2-k_F^2}{2m}$ and $k_F$ is the Fermi momentum. The 4-component Nambu spinor $\Psi=(\psi_\uparrow,\psi_\downarrow,\psi_\downarrow^\dagger,-\psi_\uparrow^\dagger)^\mathrm{T}$ is written in terms of electron annihilation (creation) operators $\psi_\sigma\,(\psi^\dagger_\sigma)$ with spin projection $\sigma$. Here $\sigma_i$ and $\tau_i$ are, respectively, Pauli matrices acting in the spin and particle-hole spaces. The spin lattice $\vec{S}_{\vec r}=\sum_j\delta(\vec r-\vec{r}_j)\vec{S}_j$ is exchange coupled to electrons with strength $J$. In what follows we consider classical spins with unit norm $|\vec{S}_j|^2=1$, while quantum spins at finite $T$ can be treated within a classical-renormalized framework \cite{Chakravarty89,Capriotti04b}.

At sufficiently weak coupling \cite{Schecter16} the electrons may be integrated out to obtain an effective adatom Heisenberg Hamiltonian
\begin{equation}\label{eq:Hspin}
H_{\mathrm{S}}=\frac{1}{2}\sum_{i\neq j}I({\bf r}_i-{\bf r}_j)\vec{S}_i\cdot\vec{S}_j.
\end{equation}
The exchange coupling $I(r)$ is given by \cite{Yao14, Schecter16}
\begin{eqnarray}
\label{eq:J-ex}
\nonumber I(r)&=&\left[\frac{v_{F}\mathrm{cos}(2k_{F}r)}{2\pi r}
+\frac{(\Delta^{2}-3\varepsilon^2)\mathrm{cos}^{2}(k_{F}r)}{2|\varepsilon|}+|\varepsilon|\right]
\\
&\times&\left(1-\frac{\varepsilon^2}{\Delta^2}\right)
\frac{e^{-2r/\xi_s}}{2(k_{F}r)^2},
\end{eqnarray}
where $v_F$ is the Fermi velocity and $\xi_s=v_F/\Delta$ $(\hbar=1)$ is the coherence length of the superconductor. In Eq.~\eqref{eq:J-ex} $\varepsilon$ is the energy of the subgap Yu-Shiba-Rusinov \cite{Luh65,Shiba68,Rusinov69} (YSR) states  formed at each magnetic adatom and is parameterized by the coupling $J$ through the relation $\varepsilon=\pm\Delta\frac{1-J^2\nu^2}{1+J^2\nu^2}$, where $\nu$ is the normal state density of states at the Fermi level.  The YSR states play an important role in enhancing the Cooper pair contribution to the indirect exchange coupling \cite{Yao14, Schecter16}, as explained below.

The first term in square brackets in Eq.~\eqref{eq:J-ex} is the standard RKKY interaction \cite{Ruderman54,Kasuya56,Yosida57}, while the remaining AFM terms arise from Cooper pairs that disfavor the pair-breaking effect of a polarized exchange field. The $\Delta^2/|\varepsilon|$ term stems from virtual Cooper pair tunneling into a pair of YSR states \cite{Yao14,Schecter16} and is valid only for $|\varepsilon|>\Delta/k_Fa$, where Cooper pairs remain off-resonant with the YSR chain. As the YSR band approaches the Fermi level ($\varepsilon\to0$), higher order spin-spin interactions become increasingly relevant. This leads to the breakdown of the Heisenberg Hamiltonian, Eq.~\eqref{eq:Hspin}, and the promotion of topological superconductivity in the YSR band \cite{Schecter16}. We will not address this interesting regime for the 2D lattice here (for the 1D case see Ref.~\cite{Schecter16}), and instead will consider the possible magnetic phases allowed by Eq.~\eqref{eq:Hspin}.

The classical spin ground state of Eq.~\eqref{eq:Hspin} is determined by the minimum Fourier component $\bf{Q}$ of the exchange interaction, $I_{{\bf q}}$. In the normal state ($\Delta=0$), one finds from Eq.~(\ref{eq:J-ex}) a FM nearest neighbor RKKY coupling in the range  $n+1/4<k_Fa/\pi< n+3/4$ with integer $n$ and an AFM nearest neighbor RKKY coupling otherwise. This generally (although not always) leads to FM ($\bf{Q}=0$) and AFM ($\bf{Q}=(\pi,\pi)$) ground states for the corresponding range of lattice constants given above.  In the case of FM order, turning on singlet superconducting correlations in the 3D electron gas generally leads to an instability towards helimagnetism due to the long-range superconducting correction in Eq.~(\ref{eq:J-ex}). For $\Delta>0$ one finds the scaling near ${\bf q}=0$: $I_{\bf q}\sim \frac{E_F}{(k_Fa)^3} (qa)^2-\frac{\Delta^2}{|\varepsilon|(k_Fa)^2}\log(qa)$, the minimization of which leads to a finite value of the ground state wavevector amplitude $Qa\sim\sqrt{\frac{\Delta a}{|\varepsilon|\xi_s}}$.
This magnetic instability is similar to the Anderson-Suhl transition in 2D and 3D spin lattices \cite{Anderson59,Aristov97} and results from the compromise between the shorter-range FM RKKY interaction and the longer-range AFM interaction mediated by Cooper pairs. Here, however, the finite codimension of the 3D superconductor with respect to the 2D spin lattice leads to a distinct scaling of $Q$ with $\Delta$ and negligible magnetic backaction on the SC order parameter \cite{Christensen16}.

The direction of wavevector ${\bf Q}$ is constrained by energetics and the symmetry of the underlying adatom lattice. We generally find that ${\bf Q}$ tends to align along the high symmetry axes for the square lattice case.  When this occurs, it implies a two-fold degenerate ground state manifold  (excluding global spin rotations) spanned by ${\bf Q}_{1,2}$, which are associated with the states  $\vec{S}_{\alpha i}=\textbf{u}\cos \textbf{Q}_\alpha\cdot\textbf{r}_i+\textbf{v}\sin \textbf{Q}_\alpha\cdot\textbf{r}_i$  \cite{Villain77}, where $\textbf{u},\,\textbf{v}$ are orthonormal vectors and $\alpha=1,2$.

We now investigate the possibility of spontaneous point group symmetry breaking for the system defined by Eq.~\eqref{eq:Hspin} using the effective exchange bond theory of Ref.~\cite{Schecter17}. This approach has the advantage of being relatively simple, and is capable of describing systems with arbitrary commensurate or incommensurate ground state wavevector ${\bf Q}_\alpha$ manifolds in the thermodynamic limit. The effective exchange bonds are defined through the spin-spin correlation function $\langle |{\vec S}_{\bf q}|^2\rangle=NT/(2K^{\rm eff}_{\bf q})$, where $N=3$ is the number of vector components of the Heisenberg spin $\vec{S}_i$ and $\langle...\rangle$ denotes a thermal average. A central result of Ref.~\cite{Schecter17} is that the exchange bonds may be determined self-consistently at leading order in $1/N$ by solving the following nonlinear bond equation
\begin{eqnarray}
\label{eq:bond}
K_\bf{q}^{\rm eff}&=& K_\bf{q}+\frac{2}{N}\int_\bf{p}\frac{1}{K^{\rm eff}_{\bf{p}+\bf{q}}}\left(\int_\bf{k}\frac{1}{K^{\rm eff}_\bf{k}}\frac{1}{K^{\rm eff}_{\bf{k}+\bf{p}}}\right)^{-1},
\end{eqnarray}
where $\int_{\bf q}$ denotes integration over the Brillouin zone, $K_{\bf q}=I_{\bf q}-I_{\bf Q}+\Omega$ and $\Omega(T)>0$ must be chosen to satisfy the sum-rule $\int_{\bf q}\frac{NT}{2K^{\rm eff}_{\bf q}}=1$, i.e. $|\vec{S}_i|^2=1$.

Point group symmetry breaking occurs through a spontaneous distortion of the effective exchange bonds, i.e. $K^{\rm eff}_{\bf q}$. In the case of broken lattice-rotation symmetry, this implies that spin correlations along orthogonal directions become distinct below a critical temperature $T_c$, e.g. $\langle\vec{S}_{i}\cdot\vec{S}_{i+x}\rangle\neq\langle\vec{S}_{i}\cdot\vec{S}_{i+y}\rangle$.
The corresponding $Z_2$ order parameters can be defined as
\begin{equation}
\label{eq:op}
\sigma^{a\{d\}}=\int_\bf{q}\langle |{\vec S}_{\bf q}|^2\rangle f_{\bf q}^{a\{d\}},
\end{equation}
where $f^{a\{d\}}_{\bf q}=(\cos q_x-\cos q_y)\{\cos(q_x+q_y)-\cos(q_x-q_y)\}$. Although $\sigma^{a\{d\}}$ both break rotation symmetry, they transform differently under mirror reflections as indicated by the form factors $f^{a\{d\}}_{\bf q}$.

%%%%%%%%%%%
%%%%%%%%%%%
\begin{figure}[b]
\includegraphics[width=\columnwidth]{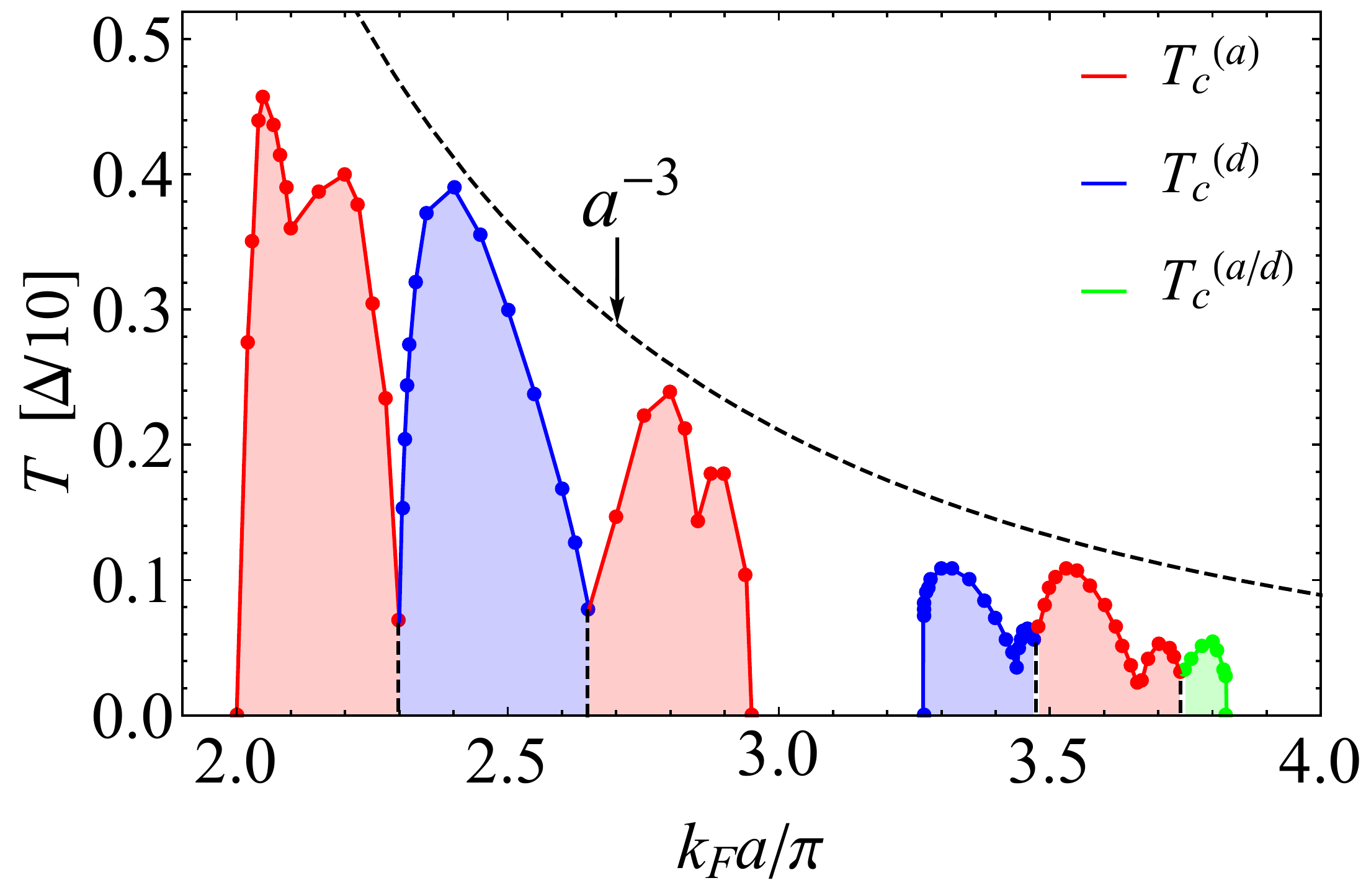}
\caption{Phase diagram of Eq.~\eqref{eq:Hspin} in the plane of temperature $T$ and adatom lattice constant $a$ for $\Delta/E_F=5\times10^{-3}$, $\varepsilon/\Delta=0.5$. The red(blue) points indicate the critical temperature for the nematic order parameter $\sigma^{a(d)}$ with axis mirror(diagonal) symmetry. The green points indicate a region of strong frustration where ${\bf Q}$ lies off the high symmetry axes, leading to nematic order without mirror symmetries, i.e. $\sigma^a\neq0,\sigma^d\neq0$.}
\label{fig:2}
\end{figure}
%%%%%%%%%%%
%%%%%%%%%%%

We solve Eq.~\eqref{eq:bond} numerically to construct the phase diagram of Eq.~\eqref{eq:Hspin}, which is presented in Fig.~\ref{fig:2}. We find broad regions of symmetry-broken phases, centered primarily around the sequence $k_Fa/\pi = n+1/2$, for integer $n$ ($n=2,3$ shown in Fig.~\ref{fig:2}), where there is a FM RKKY coupling. The overall scale of the critical temperature decreases with increasing $a$ in a power-law fashion, shown by the dashed line in Fig.~\ref{fig:2}, due to the algebraic decay of the indirect exchange coupling, Eq.~\eqref{eq:J-ex}. Near integer values of $k_Fa/\pi$ the nearest neighbor RKKY coupling is AFM, leading to regions of short-range AFM order without any symmetry breaking for all $T>0$.

Within the symmetry broken regions of Fig.~\ref{fig:2} there exists transitions of the ground state wavevector ${\bf Q}$  as a function of $k_Fa$, corresponding to a switch from axis to diagonal orientation or vice-versa \footnote{See Supplemental Material at [URL will be inserted by publisher] for some typical spin configurations in these symmetry broken regions.}. Near such points the order parameters $\sigma^a$ and $\sigma^d$ compete, leading to a sequence of bicritical points with suppressed $T_c$. For $T<T_c$ this leads to first order transitions between $\sigma^a$ and $\sigma^d$ as a function of $k_Fa$, indicated by the vertical dashed lines in Fig.~\ref{fig:2}.

%%%%%%%%%%%
%%%%%%%%%%%
\begin{figure}[b]
\includegraphics[width=\columnwidth]{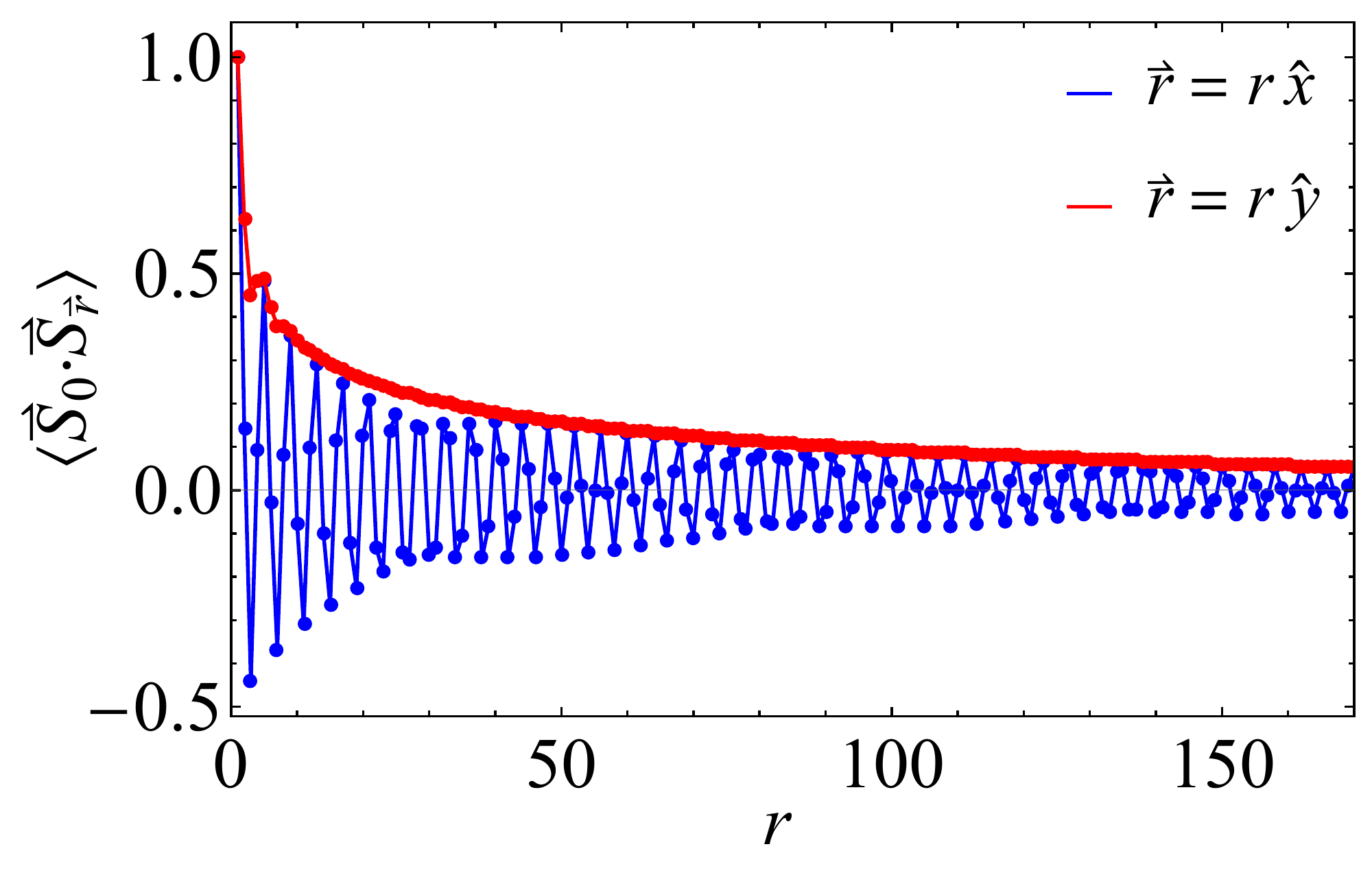}
\caption{(Color online) Real space spin-spin correlation function in the symmetry broken phase $\sigma^a\neq0$, i.e. $\langle \vec S_0\cdot \vec S_x\rangle\neq\langle \vec S_0\cdot \vec S_y\rangle$. The asymptotic form of the correlator is given by Eq.~\eqref{eq:corr2}. The parameters are $\Delta/E_F=5\times10^{-3}$, $k_Fa/\pi=3.6$, $\varepsilon/\Delta=0.5$ and $T/T_c = 0.83$, with $T_c/\Delta=8.1\times10^{-3}$. The  spin-spin correlation length is $\xi/a\approx 250$.}
\label{fig:3}
\end{figure}
%%%%%%%%%%%
%%%%%%%%%%%

The real-space spin-spin correlation function can be analyzed by Fourier transforming $NT/(2K_\bf{q}^{\rm eff})$. We show its typical spatial structure in Fig.~\ref{fig:3} in the symmetry-broken phase $T<T_c$. Apart from the anisotropic form of the correlations, $\langle\vec{S}_{i}\cdot\vec{S}_{i+x}\rangle\neq\langle\vec{S}_{i}\cdot\vec{S}_{i+y}\rangle$,  the asymptotics can be obtained by expanding $\langle|{\vec S}_{\bf q}|^2\rangle$ near ${\bf Q}_{1,2}$,
\begin{equation}
\label{eq:corr1}
\langle|{\vec S}_{{\bf q}\pm{\bf Q}_{1,2}}|^2\rangle\propto \frac{1}{{\bf q}^2+\xi^{-2}_{1,2}}.
\end{equation}
If we expand near $T_c$ we have, to linear order in $\sigma$ ($|\sigma|\ll1$), $\xi_{1,2}\approx \xi(T_c)(1\pm C \sigma)$, where $C$ is a dimensionless number that depends on the microscopic parameters.
Fourier transforming Eq.~\eqref{eq:corr1} leads to the asymptotic real-space correlation function
\begin{equation}
\label{eq:corr2}
\langle\vec S_0\cdot\vec S_{\bf r}\rangle \propto \sum_{\alpha=1,2}\frac{e^{-r/\xi_\alpha}}{\sqrt{r/\xi_\alpha}}\cos \left({\bf Q}_{\alpha}\cdot {\bf r}\right).
\end{equation}
Although  Eq.~\eqref{eq:corr2} has contributions from both ${\bf Q}_{1,2}$, only one of them is significant for $T\lesssim T_c$. This is due to the nonanalytic growth of $|\sigma|$ (and thus of $\xi_1$ or $\xi_2$) across the critical point. As a result,  Eq.~\eqref{eq:corr2} is essentially governed by a single correlation length $\xi$  both above and below $T_c$,
\begin{equation}
\label{eq:corr3}
\xi ={\rm max}\,(\xi_1,\xi_2).
\end{equation}
This behavior can be seen in Fig.~\ref{fig:3} already for $T\approx 0.8 T_c$, where $\xi/a\approx 250$.
%%%%%%%%%%%
%%%%%%%%%%%
\begin{figure}[t]
\includegraphics[width=\columnwidth]{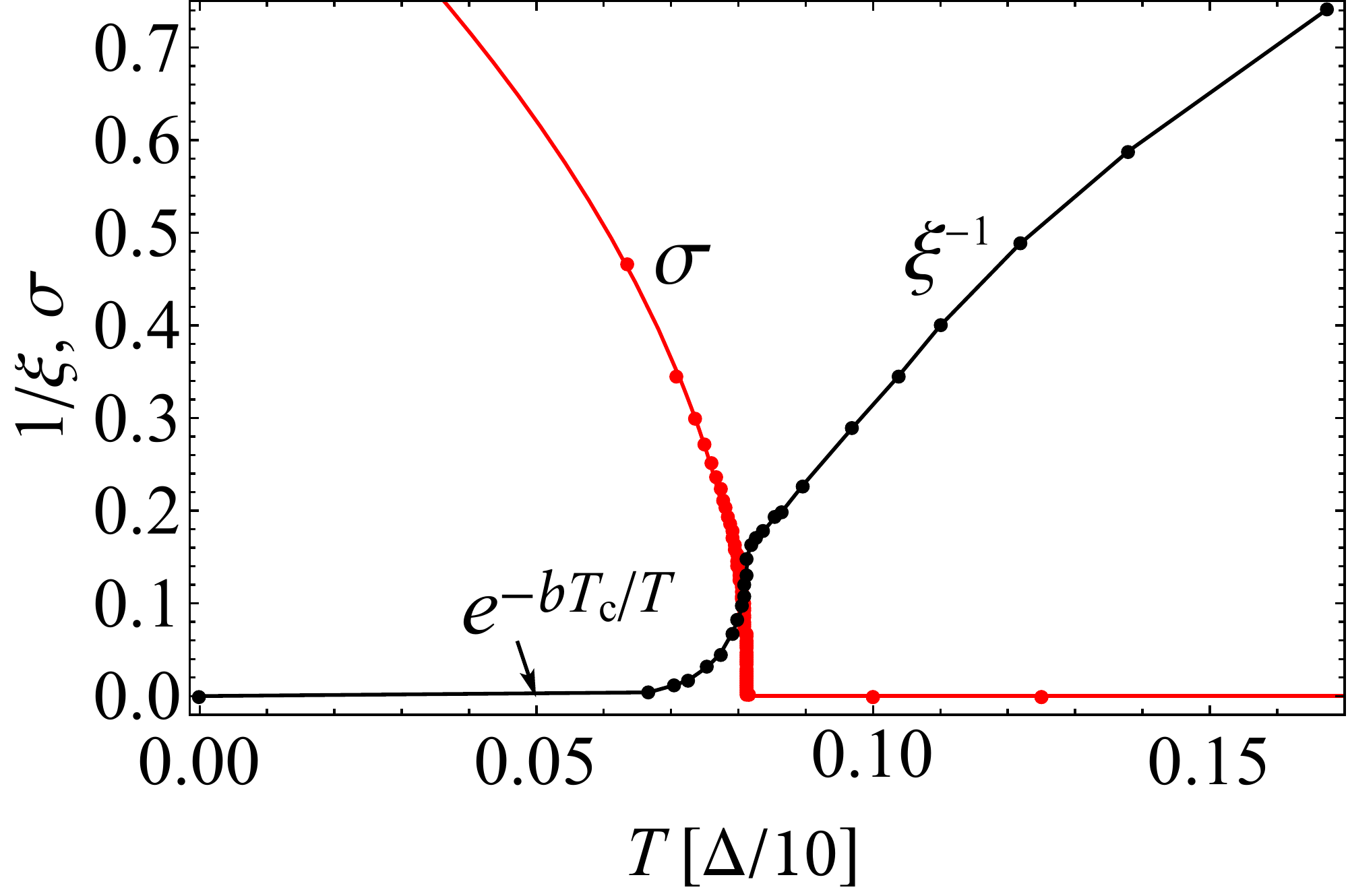}
\caption{Temperature dependence of the order parameter $\sigma=\sigma^a$ and inverse spin-spin correlation length $\xi$ for $\Delta/E_F=5\times10^{-3}$, $k_Fa/\pi=3.6$, and $\varepsilon/\Delta=0.5$. The inverse correlation length $\xi^{-1}$ decreases nonanalytically across the critical temperature $T_c/\Delta=8.1\times10^{-3}$ and decreases exponentially as $T\to 0$.}
\label{fig:4}
\end{figure}
%%%%%%%%%%%
%%%%%%%%%%%
The nonanalytic growth of $\xi$ below $T_c$ results from its relation to the order parameter, as discussed above. This is seen in Fig~\ref{fig:4} where we plot $\sigma$ and $\xi^{-1}$ as a function of $T$ in the vicinity of $T_c$. The strong decrease of $\xi^{-1}$ with the onset of $\sigma$ is evident. \begin{comment} Our numerical results from Eq.~\eqref{eq:bond} indicate a non mean-field exponent of $\beta\approx0.25$. Preliminary Monte Carlo data are not inconsistent with the expected Ising universality exponent $\beta=1/8$, but an accurate determination is challenging due to the long-range, oscillatory nature of the exchange couplings in Eq.~\eqref{eq:J-ex}. We leave this issue open for future work.\end{comment}

For $T\ll T_c$ the correlation length increases exponentially fast $\xi\propto e^{bT_c/T}$, as indicated in Fig.~\ref{fig:4}. The exponential dependence follows from the local constraint $\int_{\bf q}\langle |\vec S_{\bf q}|^2\rangle=1$, where the left side may be approximated at small $T$ as $\int_0^{1/a^2}dq^2\frac{T/(bT_c)}{q^2+\xi^{-2}}\sim T/(bT_c)\log(\xi /a)=1$, and $b$ is a parameter dependent dimensionless number.

Experimental realizations of systems described here ideally consist of thin ferromagnetic films or magnetic adatoms deposited on a 3D superconductor, with a direct exchange coupling not larger than the long-range AFM indirect exchange coupling. Since the latter scales with $\Delta$, the superconducting transition temperature sets a rough upper bound for the Curie temperature of the film. Strong magnetic anisotropy is expected to reduce the number of relevant spin components from $N=3$ to $N=2$ (easy-plane) or $N=1$ (easy-axis). Such systems may still display magnetic phases with a finite wavevector and broken point-group symmetry, but we relegate this problem to future work.

A prospective experimental system consists of magnetic manganese phthalocyanine (MnPc) molecules adsorbed on the surface of superconducting Pb~\cite{Fu07,Franke11,Fu12}. At low temperatures the MnPc molecules self-assemble into islands with square lattice symmetry and induce resolvable YSR states in the Pb substrate~\cite{Franke11}. Measurements of $a\approx 1.35\,{\rm nm}$~\cite{Fu07,Franke11} and $k_F \approx10.8\, {\rm nm}^{-1}$~\cite{Ruby16} lead to $k_Fa/\pi\approx 4.6$. Assuming $\varepsilon\approx 0.5\Delta$ with $\Delta/E_F=2(k_F\xi)^{-1}\approx2\times10^{-3}$ we find from Eq.~\eqref{eq:bond} axis-oriented nematic order below $T_c\approx 100\,{\rm mK}$. 

The wavelength $\lambda=2\pi/|{\bf Q}|$ and correlation length  associated with the short-range helimagnetic order at $T_c$ are comparable, $\xi\approx \lambda=22a\approx 30\,{\rm nm}$, with $\xi$ increasing rapidly for $T<T_c$. Due to a mismatch of lattice constants, a moir\'{e} pattern might also be detectable in the magnetic correlations. In the temperature range $T_c<T\lesssim |I_{\bf Q}|$, where $|I_{\bf Q}|\approx 560\, {\rm mK}$ is the effective Curie-Weiss temperature, we find classical helicoidal spin liquid behavior~\cite{Seabra16} associated with a ring degeneracy of $I_{\bf q}$.  In this symmetry restored regime spin correlations persist over many sites and exhibit oscillations with characteristic wavelength $\lambda$. We note that a perpendicular magnetic field $B$ could be used to tune the wavevector $|{\bf Q}|\propto \sqrt{\Delta(B)}$ and critical temperature $T_c\propto |\bf{Q}|^2\propto \Delta(B)$~\cite{Schecter17}. Interestingly, this field tuning occurs through the suppression of the gap and is distinct from tuning via the (completely screened) adatom Zeeman field.

\begin{comment}
Aside from magnetic adatoms, we speculate that thin films or strips of insulating FM EuS (bulk  $T_c\approx 17\,{\rm K}$) epitaxially deposited \cite{Katmis16} on top of a superconductor with comparable $T_c$ could conceivably be a candidate system exhibiting unconventional spin or topological superconducting orders \cite{Nakosai13}, depending e.g. on the interface contact (exchange coupling) and lattice mismatch.
\end{comment}

\acknowledgements{The Center for Quantum Devices is funded by the Danish National Research Foundation. We acknowledge support from the Villum Foundation, the Laboratory for Physical Sciences, and Microsoft (M.S.), and Grant No. 213606 from the Research Council of Norway (O.F.S.).}

\bibliography{bib-schecter17}

\begin{thebibliography}{55}%
\makeatletter
\providecommand \@ifxundefined [1]{%
 \@ifx{#1\undefined}
}%
\providecommand \@ifnum [1]{%
 \ifnum #1\expandafter \@firstoftwo
 \else \expandafter \@secondoftwo
 \fi
}%
\providecommand \@ifx [1]{%
 \ifx #1\expandafter \@firstoftwo
 \else \expandafter \@secondoftwo
 \fi
}%
\providecommand \natexlab [1]{#1}%
\providecommand \enquote  [1]{``#1''}%
\providecommand \bibnamefont  [1]{#1}%
\providecommand \bibfnamefont [1]{#1}%
\providecommand \citenamefont [1]{#1}%
\providecommand \href@noop [0]{\@secondoftwo}%
\providecommand \href [0]{\begingroup \@sanitize@url \@href}%
\providecommand \@href[1]{\@@startlink{#1}\@@href}%
\providecommand \@@href[1]{\endgroup#1\@@endlink}%
\providecommand \@sanitize@url [0]{\catcode `\\12\catcode `\$12\catcode
  `\&12\catcode `\#12\catcode `\^12\catcode `\_12\catcode `\%12\relax}%
\providecommand \@@startlink[1]{}%
\providecommand \@@endlink[0]{}%
\providecommand \url  [0]{\begingroup\@sanitize@url \@url }%
\providecommand \@url [1]{\endgroup\@href {#1}{\urlprefix }}%
\providecommand \urlprefix  [0]{URL }%
\providecommand \Eprint [0]{\href }%
\providecommand \doibase [0]{http://dx.doi.org/}%
\providecommand \selectlanguage [0]{\@gobble}%
\providecommand \bibinfo  [0]{\@secondoftwo}%
\providecommand \bibfield  [0]{\@secondoftwo}%
\providecommand \translation [1]{[#1]}%
\providecommand \BibitemOpen [0]{}%
\providecommand \bibitemStop [0]{}%
\providecommand \bibitemNoStop [0]{.\EOS\space}%
\providecommand \EOS [0]{\spacefactor3000\relax}%
\providecommand \BibitemShut  [1]{\csname bibitem#1\endcsname}%
\let\auto@bib@innerbib\@empty
%</preamble>
\bibitem [{\citenamefont {Fu}\ \emph {et~al.}(2007)\citenamefont {Fu},
  \citenamefont {Ji}, \citenamefont {Chen}, \citenamefont {Ma}, \citenamefont
  {Wu}, \citenamefont {Wang}, \citenamefont {Duan}, \citenamefont {Qiu},
  \citenamefont {Sun}, \citenamefont {Zhang}, \citenamefont {Jia},\ and\
  \citenamefont {Xue}}]{Fu07}%
  \BibitemOpen
  \bibfield  {author} {\bibinfo {author} {\bibfnamefont {Y.-S.}\ \bibnamefont
  {Fu}}, \bibinfo {author} {\bibfnamefont {S.-H.}\ \bibnamefont {Ji}}, \bibinfo
  {author} {\bibfnamefont {X.}~\bibnamefont {Chen}}, \bibinfo {author}
  {\bibfnamefont {X.-C.}\ \bibnamefont {Ma}}, \bibinfo {author} {\bibfnamefont
  {R.}~\bibnamefont {Wu}}, \bibinfo {author} {\bibfnamefont {C.-C.}\
  \bibnamefont {Wang}}, \bibinfo {author} {\bibfnamefont {W.-H.}\ \bibnamefont
  {Duan}}, \bibinfo {author} {\bibfnamefont {X.-H.}\ \bibnamefont {Qiu}},
  \bibinfo {author} {\bibfnamefont {B.}~\bibnamefont {Sun}}, \bibinfo {author}
  {\bibfnamefont {P.}~\bibnamefont {Zhang}}, \bibinfo {author} {\bibfnamefont
  {J.-F.}\ \bibnamefont {Jia}}, \ and\ \bibinfo {author} {\bibfnamefont
  {Q.-K.}\ \bibnamefont {Xue}},\ }\href {\doibase
  10.1103/PhysRevLett.99.256601} {\bibfield  {journal} {\bibinfo  {journal}
  {Phys. Rev. Lett.}\ }\textbf {\bibinfo {volume} {99}},\ \bibinfo {pages}
  {256601} (\bibinfo {year} {2007})}\BibitemShut {NoStop}%
\bibitem [{\citenamefont {Ji}\ \emph {et~al.}(2008)\citenamefont {Ji},
  \citenamefont {Zhang}, \citenamefont {Fu}, \citenamefont {Chen},
  \citenamefont {Ma}, \citenamefont {Li}, \citenamefont {Duan}, \citenamefont
  {Jia},\ and\ \citenamefont {Xue}}]{Ji08}%
  \BibitemOpen
  \bibfield  {author} {\bibinfo {author} {\bibfnamefont {S.-H.}\ \bibnamefont
  {Ji}}, \bibinfo {author} {\bibfnamefont {T.}~\bibnamefont {Zhang}}, \bibinfo
  {author} {\bibfnamefont {Y.-S.}\ \bibnamefont {Fu}}, \bibinfo {author}
  {\bibfnamefont {X.}~\bibnamefont {Chen}}, \bibinfo {author} {\bibfnamefont
  {X.-C.}\ \bibnamefont {Ma}}, \bibinfo {author} {\bibfnamefont
  {J.}~\bibnamefont {Li}}, \bibinfo {author} {\bibfnamefont {W.-H.}\
  \bibnamefont {Duan}}, \bibinfo {author} {\bibfnamefont {J.-F.}\ \bibnamefont
  {Jia}}, \ and\ \bibinfo {author} {\bibfnamefont {Q.-K.}\ \bibnamefont
  {Xue}},\ }\href {\doibase 10.1103/PhysRevLett.100.226801} {\bibfield
  {journal} {\bibinfo  {journal} {Phys. Rev. Lett.}\ }\textbf {\bibinfo
  {volume} {100}},\ \bibinfo {pages} {226801} (\bibinfo {year}
  {2008})}\BibitemShut {NoStop}%
\bibitem [{\citenamefont {Franke}\ \emph {et~al.}(2011)\citenamefont {Franke},
  \citenamefont {Schulze},\ and\ \citenamefont {Pascual}}]{Franke11}%
  \BibitemOpen
  \bibfield  {author} {\bibinfo {author} {\bibfnamefont {K.~J.}\ \bibnamefont
  {Franke}}, \bibinfo {author} {\bibfnamefont {G.}~\bibnamefont {Schulze}}, \
  and\ \bibinfo {author} {\bibfnamefont {J.~I.}\ \bibnamefont {Pascual}},\
  }\href {\doibase 10.1126/science.1202204} {\bibfield  {journal} {\bibinfo
  {journal} {Science}\ }\textbf {\bibinfo {volume} {332}},\ \bibinfo {pages}
  {940} (\bibinfo {year} {2011})}\BibitemShut {NoStop}%
\bibitem [{\citenamefont {Fu}\ \emph {et~al.}(2012)\citenamefont {Fu},
  \citenamefont {Xue},\ and\ \citenamefont {Wiesendanger}}]{Fu12}%
  \BibitemOpen
  \bibfield  {author} {\bibinfo {author} {\bibfnamefont {Y.-S.}\ \bibnamefont
  {Fu}}, \bibinfo {author} {\bibfnamefont {Q.-K.}\ \bibnamefont {Xue}}, \ and\
  \bibinfo {author} {\bibfnamefont {R.}~\bibnamefont {Wiesendanger}},\ }\href
  {\doibase 10.1103/PhysRevLett.108.087203} {\bibfield  {journal} {\bibinfo
  {journal} {Phys. Rev. Lett.}\ }\textbf {\bibinfo {volume} {108}},\ \bibinfo
  {pages} {087203} (\bibinfo {year} {2012})}\BibitemShut {NoStop}%
\bibitem [{\citenamefont {Nadj-Perge}\ \emph {et~al.}(2014)\citenamefont
  {Nadj-Perge}, \citenamefont {Drozdov}, \citenamefont {Li}, \citenamefont
  {Chen}, \citenamefont {Jeon}, \citenamefont {Seo}, \citenamefont {MacDonald},
  \citenamefont {Bernevig},\ and\ \citenamefont {Yazdani}}]{Nadj-Perge14}%
  \BibitemOpen
  \bibfield  {author} {\bibinfo {author} {\bibfnamefont {S.}~\bibnamefont
  {Nadj-Perge}}, \bibinfo {author} {\bibfnamefont {I.~K.}\ \bibnamefont
  {Drozdov}}, \bibinfo {author} {\bibfnamefont {J.}~\bibnamefont {Li}},
  \bibinfo {author} {\bibfnamefont {H.}~\bibnamefont {Chen}}, \bibinfo {author}
  {\bibfnamefont {S.}~\bibnamefont {Jeon}}, \bibinfo {author} {\bibfnamefont
  {J.}~\bibnamefont {Seo}}, \bibinfo {author} {\bibfnamefont {A.~H.}\
  \bibnamefont {MacDonald}}, \bibinfo {author} {\bibfnamefont {B.~A.}\
  \bibnamefont {Bernevig}}, \ and\ \bibinfo {author} {\bibfnamefont
  {A.}~\bibnamefont {Yazdani}},\ }\href {\doibase 10.1126/science.1259327}
  {\bibfield  {journal} {\bibinfo  {journal} {Science}\ }\textbf {\bibinfo
  {volume} {346}},\ \bibinfo {pages} {602} (\bibinfo {year}
  {2014})}\BibitemShut {NoStop}%
\bibitem [{\citenamefont {Hatter}\ \emph {et~al.}(2015)\citenamefont {Hatter},
  \citenamefont {Heinrich}, \citenamefont {Ruby}, \citenamefont {Pascual},\
  and\ \citenamefont {Franke}}]{Hatter15}%
  \BibitemOpen
  \bibfield  {author} {\bibinfo {author} {\bibfnamefont {N.}~\bibnamefont
  {Hatter}}, \bibinfo {author} {\bibfnamefont {B.~W.}\ \bibnamefont
  {Heinrich}}, \bibinfo {author} {\bibfnamefont {M.}~\bibnamefont {Ruby}},
  \bibinfo {author} {\bibfnamefont {J.~I.}\ \bibnamefont {Pascual}}, \ and\
  \bibinfo {author} {\bibfnamefont {K.~J.}\ \bibnamefont {Franke}},\ }\href
  {\doibase doi:10.1038/ncomms9988} {\bibfield  {journal} {\bibinfo  {journal}
  {Nature Communications}\ }\textbf {\bibinfo {volume} {6}},\ \bibinfo {pages}
  {8988} (\bibinfo {year} {2015})}\BibitemShut {NoStop}%
\bibitem [{\citenamefont {Ruby}\ \emph {et~al.}(2015)\citenamefont {Ruby},
  \citenamefont {Pientka}, \citenamefont {Peng}, \citenamefont {von Oppen},
  \citenamefont {Heinrich},\ and\ \citenamefont {Franke}}]{Ruby15}%
  \BibitemOpen
  \bibfield  {author} {\bibinfo {author} {\bibfnamefont {M.}~\bibnamefont
  {Ruby}}, \bibinfo {author} {\bibfnamefont {F.}~\bibnamefont {Pientka}},
  \bibinfo {author} {\bibfnamefont {Y.}~\bibnamefont {Peng}}, \bibinfo {author}
  {\bibfnamefont {F.}~\bibnamefont {von Oppen}}, \bibinfo {author}
  {\bibfnamefont {B.~W.}\ \bibnamefont {Heinrich}}, \ and\ \bibinfo {author}
  {\bibfnamefont {K.~J.}\ \bibnamefont {Franke}},\ }\href {\doibase
  10.1103/PhysRevLett.115.197204} {\bibfield  {journal} {\bibinfo  {journal}
  {Phys. Rev. Lett.}\ }\textbf {\bibinfo {volume} {115}},\ \bibinfo {pages}
  {197204} (\bibinfo {year} {2015})}\BibitemShut {NoStop}%
\bibitem [{\citenamefont {Ruby}\ \emph {et~al.}(2016)\citenamefont {Ruby},
  \citenamefont {Peng}, \citenamefont {von Oppen}, \citenamefont {Heinrich},\
  and\ \citenamefont {Franke}}]{Ruby16}%
  \BibitemOpen
  \bibfield  {author} {\bibinfo {author} {\bibfnamefont {M.}~\bibnamefont
  {Ruby}}, \bibinfo {author} {\bibfnamefont {Y.}~\bibnamefont {Peng}}, \bibinfo
  {author} {\bibfnamefont {F.}~\bibnamefont {von Oppen}}, \bibinfo {author}
  {\bibfnamefont {B.~W.}\ \bibnamefont {Heinrich}}, \ and\ \bibinfo {author}
  {\bibfnamefont {K.~J.}\ \bibnamefont {Franke}},\ }\href {\doibase
  10.1103/PhysRevLett.117.186801} {\bibfield  {journal} {\bibinfo  {journal}
  {Phys. Rev. Lett.}\ }\textbf {\bibinfo {volume} {117}},\ \bibinfo {pages}
  {186801} (\bibinfo {year} {2016})}\BibitemShut {NoStop}%
\bibitem [{\citenamefont {Pawlak}\ \emph {et~al.}(2016)\citenamefont {Pawlak},
  \citenamefont {Kisiel}, \citenamefont {Klinovaja}, \citenamefont {Meier},
  \citenamefont {Kawai}, \citenamefont {Glatzel}, \citenamefont {Loss},\ and\
  \citenamefont {Meyer}}]{Pawlak16}%
  \BibitemOpen
  \bibfield  {author} {\bibinfo {author} {\bibfnamefont {R.}~\bibnamefont
  {Pawlak}}, \bibinfo {author} {\bibfnamefont {M.}~\bibnamefont {Kisiel}},
  \bibinfo {author} {\bibfnamefont {J.}~\bibnamefont {Klinovaja}}, \bibinfo
  {author} {\bibfnamefont {T.}~\bibnamefont {Meier}}, \bibinfo {author}
  {\bibfnamefont {S.}~\bibnamefont {Kawai}}, \bibinfo {author} {\bibfnamefont
  {T.}~\bibnamefont {Glatzel}}, \bibinfo {author} {\bibfnamefont
  {D.}~\bibnamefont {Loss}}, \ and\ \bibinfo {author} {\bibfnamefont
  {E.}~\bibnamefont {Meyer}},\ }\href {\doibase doi:10.1038/npjqi.2016.35}
  {\bibfield  {journal} {\bibinfo  {journal} {npj Quantum Information}\
  }\textbf {\bibinfo {volume} {2}},\ \bibinfo {pages} {16035} (\bibinfo {year}
  {2016})}\BibitemShut {NoStop}%
\bibitem [{\citenamefont {Heinrich}\ \emph {et~al.}(2017)\citenamefont
  {Heinrich}, \citenamefont {Pascual},\ and\ \citenamefont
  {Franke}}]{Heinrich17}%
  \BibitemOpen
  \bibfield  {author} {\bibinfo {author} {\bibfnamefont {B.~W.}\ \bibnamefont
  {Heinrich}}, \bibinfo {author} {\bibfnamefont {J.~I.}\ \bibnamefont
  {Pascual}}, \ and\ \bibinfo {author} {\bibfnamefont {K.~J.}\ \bibnamefont
  {Franke}},\ }\href@noop {} {\bibfield  {journal} {\bibinfo  {journal}
  {arXiv:1705.03672v1}\ } (\bibinfo {year} {2017})}\BibitemShut {NoStop}%
\bibitem [{\citenamefont {Kezilebieke}\ \emph {et~al.}(2017)\citenamefont
  {Kezilebieke}, \citenamefont {Dvorak}, \citenamefont {Ojanen},\ and\
  \citenamefont {Liljeroth}}]{Kezilebieke17}%
  \BibitemOpen
  \bibfield  {author} {\bibinfo {author} {\bibfnamefont {S.}~\bibnamefont
  {Kezilebieke}}, \bibinfo {author} {\bibfnamefont {M.}~\bibnamefont {Dvorak}},
  \bibinfo {author} {\bibfnamefont {T.}~\bibnamefont {Ojanen}}, \ and\ \bibinfo
  {author} {\bibfnamefont {P.}~\bibnamefont {Liljeroth}},\ }\href@noop {}
  {\bibfield  {journal} {\bibinfo  {journal} {arXiv:1701.03288}\ } (\bibinfo
  {year} {2017})}\BibitemShut {NoStop}%
\bibitem [{\citenamefont {Ruby}\ \emph {et~al.}(2017)\citenamefont {Ruby},
  \citenamefont {Heinrich}, \citenamefont {Peng}, \citenamefont {von Oppen},\
  and\ \citenamefont {Franke}}]{Ruby17}%
  \BibitemOpen
  \bibfield  {author} {\bibinfo {author} {\bibfnamefont {M.}~\bibnamefont
  {Ruby}}, \bibinfo {author} {\bibfnamefont {B.~W.}\ \bibnamefont {Heinrich}},
  \bibinfo {author} {\bibfnamefont {Y.}~\bibnamefont {Peng}}, \bibinfo {author}
  {\bibfnamefont {F.}~\bibnamefont {von Oppen}}, \ and\ \bibinfo {author}
  {\bibfnamefont {K.~J.}\ \bibnamefont {Franke}},\ }\href {\doibase
  10.1021/acs.nanolett.7b01728} {\bibfield  {journal} {\bibinfo  {journal}
  {Nano Letters}\ }\textbf {\bibinfo {volume} {17}},\ \bibinfo {pages} {4473}
  (\bibinfo {year} {2017})}\BibitemShut {NoStop}%
\bibitem [{\citenamefont {Choi~et al.}(2017)}]{Choi17}%
  \BibitemOpen
  \bibfield  {author} {\bibinfo {author} {\bibfnamefont {D.-J.}\ \bibnamefont
  {Choi~et al.}},\ }\href@noop {} {\bibfield  {journal} {\bibinfo  {journal}
  {arXiv:1709.09224v1}\ } (\bibinfo {year} {2017})}\BibitemShut {NoStop}%
\bibitem [{\citenamefont {Hatter}\ \emph {et~al.}(2017)\citenamefont {Hatter},
  \citenamefont {Heinrich}, \citenamefont {Rolf},\ and\ \citenamefont
  {Franke}}]{Hatter17}%
  \BibitemOpen
  \bibfield  {author} {\bibinfo {author} {\bibfnamefont {N.}~\bibnamefont
  {Hatter}}, \bibinfo {author} {\bibfnamefont {B.~W.}\ \bibnamefont
  {Heinrich}}, \bibinfo {author} {\bibfnamefont {H.}~\bibnamefont {Rolf}}, \
  and\ \bibinfo {author} {\bibfnamefont {K.}~\bibnamefont {Franke}},\
  }\href@noop {} {\bibfield  {journal} {\bibinfo  {journal} {arXiv:1710.04599}\
  } (\bibinfo {year} {2017})}\BibitemShut {NoStop}%
\bibitem [{\citenamefont {Pientka}\ \emph {et~al.}(2013)\citenamefont
  {Pientka}, \citenamefont {Glazman},\ and\ \citenamefont {von
  Oppen}}]{Pientka13}%
  \BibitemOpen
  \bibfield  {author} {\bibinfo {author} {\bibfnamefont {F.}~\bibnamefont
  {Pientka}}, \bibinfo {author} {\bibfnamefont {L.~I.}\ \bibnamefont
  {Glazman}}, \ and\ \bibinfo {author} {\bibfnamefont {F.}~\bibnamefont {von
  Oppen}},\ }\href {\doibase 10.1103/PhysRevB.88.155420} {\bibfield  {journal}
  {\bibinfo  {journal} {Phys. Rev. B}\ }\textbf {\bibinfo {volume} {88}},\
  \bibinfo {pages} {155420} (\bibinfo {year} {2013})}\BibitemShut {NoStop}%
\bibitem [{\citenamefont {Pientka}\ \emph {et~al.}(2014)\citenamefont
  {Pientka}, \citenamefont {Glazman},\ and\ \citenamefont {von
  Oppen}}]{Pientka14}%
  \BibitemOpen
  \bibfield  {author} {\bibinfo {author} {\bibfnamefont {F.}~\bibnamefont
  {Pientka}}, \bibinfo {author} {\bibfnamefont {L.~I.}\ \bibnamefont
  {Glazman}}, \ and\ \bibinfo {author} {\bibfnamefont {F.}~\bibnamefont {von
  Oppen}},\ }\href {\doibase 10.1103/PhysRevB.89.180505} {\bibfield  {journal}
  {\bibinfo  {journal} {Phys. Rev. B}\ }\textbf {\bibinfo {volume} {89}},\
  \bibinfo {pages} {180505} (\bibinfo {year} {2014})}\BibitemShut {NoStop}%
\bibitem [{\citenamefont {Peng}\ \emph {et~al.}(2015)\citenamefont {Peng},
  \citenamefont {Pientka}, \citenamefont {Glazman},\ and\ \citenamefont {von
  Oppen}}]{Peng15}%
  \BibitemOpen
  \bibfield  {author} {\bibinfo {author} {\bibfnamefont {Y.}~\bibnamefont
  {Peng}}, \bibinfo {author} {\bibfnamefont {F.}~\bibnamefont {Pientka}},
  \bibinfo {author} {\bibfnamefont {L.~I.}\ \bibnamefont {Glazman}}, \ and\
  \bibinfo {author} {\bibfnamefont {F.}~\bibnamefont {von Oppen}},\ }\href
  {\doibase 10.1103/PhysRevLett.114.106801} {\bibfield  {journal} {\bibinfo
  {journal} {Phys. Rev. Lett.}\ }\textbf {\bibinfo {volume} {114}},\ \bibinfo
  {pages} {106801} (\bibinfo {year} {2015})}\BibitemShut {NoStop}%
\bibitem [{\citenamefont {Klinovaja}\ \emph {et~al.}(2013)\citenamefont
  {Klinovaja}, \citenamefont {Stano}, \citenamefont {Yazdani},\ and\
  \citenamefont {Loss}}]{Klinovaja13}%
  \BibitemOpen
  \bibfield  {author} {\bibinfo {author} {\bibfnamefont {J.}~\bibnamefont
  {Klinovaja}}, \bibinfo {author} {\bibfnamefont {P.}~\bibnamefont {Stano}},
  \bibinfo {author} {\bibfnamefont {A.}~\bibnamefont {Yazdani}}, \ and\
  \bibinfo {author} {\bibfnamefont {D.}~\bibnamefont {Loss}},\ }\href {\doibase
  10.1103/PhysRevLett.111.186805} {\bibfield  {journal} {\bibinfo  {journal}
  {Phys. Rev. Lett.}\ }\textbf {\bibinfo {volume} {111}},\ \bibinfo {pages}
  {186805} (\bibinfo {year} {2013})}\BibitemShut {NoStop}%
\bibitem [{\citenamefont {Braunecker}\ and\ \citenamefont
  {Simon}(2013)}]{Braunecker13}%
  \BibitemOpen
  \bibfield  {author} {\bibinfo {author} {\bibfnamefont {B.}~\bibnamefont
  {Braunecker}}\ and\ \bibinfo {author} {\bibfnamefont {P.}~\bibnamefont
  {Simon}},\ }\href {\doibase 10.1103/PhysRevLett.111.147202} {\bibfield
  {journal} {\bibinfo  {journal} {Phys. Rev. Lett.}\ }\textbf {\bibinfo
  {volume} {111}},\ \bibinfo {pages} {147202} (\bibinfo {year}
  {2013})}\BibitemShut {NoStop}%
\bibitem [{\citenamefont {Vazifeh}\ and\ \citenamefont
  {Franz}(2013)}]{Vazifeh13}%
  \BibitemOpen
  \bibfield  {author} {\bibinfo {author} {\bibfnamefont {M.~M.}\ \bibnamefont
  {Vazifeh}}\ and\ \bibinfo {author} {\bibfnamefont {M.}~\bibnamefont
  {Franz}},\ }\href {\doibase 10.1103/PhysRevLett.111.206802} {\bibfield
  {journal} {\bibinfo  {journal} {Phys. Rev. Lett.}\ }\textbf {\bibinfo
  {volume} {111}},\ \bibinfo {pages} {206802} (\bibinfo {year}
  {2013})}\BibitemShut {NoStop}%
\bibitem [{\citenamefont {Kim}\ \emph {et~al.}(2014)\citenamefont {Kim},
  \citenamefont {Cheng}, \citenamefont {Bauer}, \citenamefont {Lutchyn},\ and\
  \citenamefont {Das~Sarma}}]{Kim14}%
  \BibitemOpen
  \bibfield  {author} {\bibinfo {author} {\bibfnamefont {Y.}~\bibnamefont
  {Kim}}, \bibinfo {author} {\bibfnamefont {M.}~\bibnamefont {Cheng}}, \bibinfo
  {author} {\bibfnamefont {B.}~\bibnamefont {Bauer}}, \bibinfo {author}
  {\bibfnamefont {R.~M.}\ \bibnamefont {Lutchyn}}, \ and\ \bibinfo {author}
  {\bibfnamefont {S.}~\bibnamefont {Das~Sarma}},\ }\href {\doibase
  10.1103/PhysRevB.90.060401} {\bibfield  {journal} {\bibinfo  {journal} {Phys.
  Rev. B}\ }\textbf {\bibinfo {volume} {90}},\ \bibinfo {pages} {060401}
  (\bibinfo {year} {2014})}\BibitemShut {NoStop}%
\bibitem [{\citenamefont {Heimes}\ \emph {et~al.}(2014)\citenamefont {Heimes},
  \citenamefont {Kotetes},\ and\ \citenamefont {Sch\"on}}]{Heimes14}%
  \BibitemOpen
  \bibfield  {author} {\bibinfo {author} {\bibfnamefont {A.}~\bibnamefont
  {Heimes}}, \bibinfo {author} {\bibfnamefont {P.}~\bibnamefont {Kotetes}}, \
  and\ \bibinfo {author} {\bibfnamefont {G.}~\bibnamefont {Sch\"on}},\ }\href
  {\doibase 10.1103/PhysRevB.90.060507} {\bibfield  {journal} {\bibinfo
  {journal} {Phys. Rev. B}\ }\textbf {\bibinfo {volume} {90}},\ \bibinfo
  {pages} {060507} (\bibinfo {year} {2014})}\BibitemShut {NoStop}%
\bibitem [{\citenamefont {Heimes}\ \emph {et~al.}(2015)\citenamefont {Heimes},
  \citenamefont {Mendler},\ and\ \citenamefont {Kotetes}}]{Heimes15}%
  \BibitemOpen
  \bibfield  {author} {\bibinfo {author} {\bibfnamefont {A.}~\bibnamefont
  {Heimes}}, \bibinfo {author} {\bibfnamefont {D.}~\bibnamefont {Mendler}}, \
  and\ \bibinfo {author} {\bibfnamefont {P.}~\bibnamefont {Kotetes}},\ }\href
  {http://stacks.iop.org/1367-2630/17/i=2/a=023051} {\bibfield  {journal}
  {\bibinfo  {journal} {New Journal of Physics}\ }\textbf {\bibinfo {volume}
  {17}},\ \bibinfo {pages} {023051} (\bibinfo {year} {2015})}\BibitemShut
  {NoStop}%
\bibitem [{\citenamefont {Reis}\ \emph {et~al.}(2014)\citenamefont {Reis},
  \citenamefont {Marchand},\ and\ \citenamefont {Franz}}]{Reis14}%
  \BibitemOpen
  \bibfield  {author} {\bibinfo {author} {\bibfnamefont {I.}~\bibnamefont
  {Reis}}, \bibinfo {author} {\bibfnamefont {D.~J.~J.}\ \bibnamefont
  {Marchand}}, \ and\ \bibinfo {author} {\bibfnamefont {M.}~\bibnamefont
  {Franz}},\ }\href {\doibase 10.1103/PhysRevB.90.085124} {\bibfield  {journal}
  {\bibinfo  {journal} {Phys. Rev. B}\ }\textbf {\bibinfo {volume} {90}},\
  \bibinfo {pages} {085124} (\bibinfo {year} {2014})}\BibitemShut {NoStop}%
\bibitem [{\citenamefont {Brydon}\ \emph {et~al.}(2015)\citenamefont {Brydon},
  \citenamefont {Das~Sarma}, \citenamefont {Hui},\ and\ \citenamefont
  {Sau}}]{Brydon15}%
  \BibitemOpen
  \bibfield  {author} {\bibinfo {author} {\bibfnamefont {P.~M.~R.}\
  \bibnamefont {Brydon}}, \bibinfo {author} {\bibfnamefont {S.}~\bibnamefont
  {Das~Sarma}}, \bibinfo {author} {\bibfnamefont {H.-Y.}\ \bibnamefont {Hui}},
  \ and\ \bibinfo {author} {\bibfnamefont {J.~D.}\ \bibnamefont {Sau}},\ }\href
  {\doibase 10.1103/PhysRevB.91.064505} {\bibfield  {journal} {\bibinfo
  {journal} {Phys. Rev. B}\ }\textbf {\bibinfo {volume} {91}},\ \bibinfo
  {pages} {064505} (\bibinfo {year} {2015})}\BibitemShut {NoStop}%
\bibitem [{\citenamefont {Li}\ \emph {et~al.}(2014)\citenamefont {Li},
  \citenamefont {Chen}, \citenamefont {Drozdov}, \citenamefont {Yazdani},
  \citenamefont {Bernevig},\ and\ \citenamefont {MacDonald}}]{Li14}%
  \BibitemOpen
  \bibfield  {author} {\bibinfo {author} {\bibfnamefont {J.}~\bibnamefont
  {Li}}, \bibinfo {author} {\bibfnamefont {H.}~\bibnamefont {Chen}}, \bibinfo
  {author} {\bibfnamefont {I.~K.}\ \bibnamefont {Drozdov}}, \bibinfo {author}
  {\bibfnamefont {A.}~\bibnamefont {Yazdani}}, \bibinfo {author} {\bibfnamefont
  {B.~A.}\ \bibnamefont {Bernevig}}, \ and\ \bibinfo {author} {\bibfnamefont
  {A.~H.}\ \bibnamefont {MacDonald}},\ }\href {\doibase
  10.1103/PhysRevB.90.235433} {\bibfield  {journal} {\bibinfo  {journal} {Phys.
  Rev. B}\ }\textbf {\bibinfo {volume} {90}},\ \bibinfo {pages} {235433}
  (\bibinfo {year} {2014})}\BibitemShut {NoStop}%
\bibitem [{\citenamefont {Schecter}\ \emph {et~al.}(2015)\citenamefont
  {Schecter}, \citenamefont {Rudner},\ and\ \citenamefont
  {Flensberg}}]{Schecter15}%
  \BibitemOpen
  \bibfield  {author} {\bibinfo {author} {\bibfnamefont {M.}~\bibnamefont
  {Schecter}}, \bibinfo {author} {\bibfnamefont {M.~S.}\ \bibnamefont
  {Rudner}}, \ and\ \bibinfo {author} {\bibfnamefont {K.}~\bibnamefont
  {Flensberg}},\ }\href {\doibase 10.1103/PhysRevLett.114.247205} {\bibfield
  {journal} {\bibinfo  {journal} {Phys. Rev. Lett.}\ }\textbf {\bibinfo
  {volume} {114}},\ \bibinfo {pages} {247205} (\bibinfo {year}
  {2015})}\BibitemShut {NoStop}%
\bibitem [{\citenamefont {Choy}\ \emph {et~al.}(2011)\citenamefont {Choy},
  \citenamefont {Edge}, \citenamefont {Akhmerov},\ and\ \citenamefont
  {Beenakker}}]{Choy11}%
  \BibitemOpen
  \bibfield  {author} {\bibinfo {author} {\bibfnamefont {T.-P.}\ \bibnamefont
  {Choy}}, \bibinfo {author} {\bibfnamefont {J.~M.}\ \bibnamefont {Edge}},
  \bibinfo {author} {\bibfnamefont {A.~R.}\ \bibnamefont {Akhmerov}}, \ and\
  \bibinfo {author} {\bibfnamefont {C.~W.~J.}\ \bibnamefont {Beenakker}},\
  }\href {\doibase 10.1103/PhysRevB.84.195442} {\bibfield  {journal} {\bibinfo
  {journal} {Phys. Rev. B}\ }\textbf {\bibinfo {volume} {84}},\ \bibinfo
  {pages} {195442} (\bibinfo {year} {2011})}\BibitemShut {NoStop}%
\bibitem [{\citenamefont {Martin}\ and\ \citenamefont
  {Morpurgo}(2012)}]{Martin12}%
  \BibitemOpen
  \bibfield  {author} {\bibinfo {author} {\bibfnamefont {I.}~\bibnamefont
  {Martin}}\ and\ \bibinfo {author} {\bibfnamefont {A.~F.}\ \bibnamefont
  {Morpurgo}},\ }\href {\doibase 10.1103/PhysRevB.85.144505} {\bibfield
  {journal} {\bibinfo  {journal} {Phys. Rev. B}\ }\textbf {\bibinfo {volume}
  {85}},\ \bibinfo {pages} {144505} (\bibinfo {year} {2012})}\BibitemShut
  {NoStop}%
\bibitem [{\citenamefont {Nakosai}\ \emph {et~al.}(2013)\citenamefont
  {Nakosai}, \citenamefont {Tanaka},\ and\ \citenamefont
  {Nagaosa}}]{Nakosai13}%
  \BibitemOpen
  \bibfield  {author} {\bibinfo {author} {\bibfnamefont {S.}~\bibnamefont
  {Nakosai}}, \bibinfo {author} {\bibfnamefont {Y.}~\bibnamefont {Tanaka}}, \
  and\ \bibinfo {author} {\bibfnamefont {N.}~\bibnamefont {Nagaosa}},\ }\href
  {\doibase 10.1103/PhysRevB.88.180503} {\bibfield  {journal} {\bibinfo
  {journal} {Phys. Rev. B}\ }\textbf {\bibinfo {volume} {88}},\ \bibinfo
  {pages} {180503} (\bibinfo {year} {2013})}\BibitemShut {NoStop}%
\bibitem [{\citenamefont {R\"ontynen}\ and\ \citenamefont
  {Ojanen}(2015)}]{Rontynen15}%
  \BibitemOpen
  \bibfield  {author} {\bibinfo {author} {\bibfnamefont {J.}~\bibnamefont
  {R\"ontynen}}\ and\ \bibinfo {author} {\bibfnamefont {T.}~\bibnamefont
  {Ojanen}},\ }\href {\doibase 10.1103/PhysRevLett.114.236803} {\bibfield
  {journal} {\bibinfo  {journal} {Phys. Rev. Lett.}\ }\textbf {\bibinfo
  {volume} {114}},\ \bibinfo {pages} {236803} (\bibinfo {year}
  {2015})}\BibitemShut {NoStop}%
\bibitem [{\citenamefont {R\"ontynen}\ and\ \citenamefont
  {Ojanen}(2016)}]{Rontynen16}%
  \BibitemOpen
  \bibfield  {author} {\bibinfo {author} {\bibfnamefont {J.}~\bibnamefont
  {R\"ontynen}}\ and\ \bibinfo {author} {\bibfnamefont {T.}~\bibnamefont
  {Ojanen}},\ }\href {\doibase 10.1103/PhysRevB.93.094521} {\bibfield
  {journal} {\bibinfo  {journal} {Phys. Rev. B}\ }\textbf {\bibinfo {volume}
  {93}},\ \bibinfo {pages} {094521} (\bibinfo {year} {2016})}\BibitemShut
  {NoStop}%
\bibitem [{\citenamefont {Ruderman}\ and\ \citenamefont
  {Kittel}(1954)}]{Ruderman54}%
  \BibitemOpen
  \bibfield  {author} {\bibinfo {author} {\bibfnamefont {M.~A.}\ \bibnamefont
  {Ruderman}}\ and\ \bibinfo {author} {\bibfnamefont {C.}~\bibnamefont
  {Kittel}},\ }\href {\doibase 10.1103/PhysRev.96.99} {\bibfield  {journal}
  {\bibinfo  {journal} {Phys. Rev.}\ }\textbf {\bibinfo {volume} {96}},\
  \bibinfo {pages} {99} (\bibinfo {year} {1954})}\BibitemShut {NoStop}%
\bibitem [{\citenamefont {Kasuya}(1956)}]{Kasuya56}%
  \BibitemOpen
  \bibfield  {author} {\bibinfo {author} {\bibfnamefont {T.}~\bibnamefont
  {Kasuya}},\ }\href {\doibase 10.1143/PTP.16.45} {\bibfield  {journal}
  {\bibinfo  {journal} {Progress of Theoretical Physics}\ }\textbf {\bibinfo
  {volume} {16}},\ \bibinfo {pages} {45} (\bibinfo {year} {1956})}\BibitemShut
  {NoStop}%
\bibitem [{\citenamefont {Yosida}(1957)}]{Yosida57}%
  \BibitemOpen
  \bibfield  {author} {\bibinfo {author} {\bibfnamefont {K.}~\bibnamefont
  {Yosida}},\ }\href {\doibase 10.1103/PhysRev.106.893} {\bibfield  {journal}
  {\bibinfo  {journal} {Phys. Rev.}\ }\textbf {\bibinfo {volume} {106}},\
  \bibinfo {pages} {893} (\bibinfo {year} {1957})}\BibitemShut {NoStop}%
\bibitem [{\citenamefont {Abrikosov}(1988)}]{Abrikosov88}%
  \BibitemOpen
  \bibfield  {author} {\bibinfo {author} {\bibfnamefont {A.~A.}\ \bibnamefont
  {Abrikosov}},\ }\href@noop {} {\emph {\bibinfo {title} {Fundamentals of the
  Theory of Metals}}}\ (\bibinfo  {publisher} {North-Holland},\ \bibinfo {year}
  {1988})\BibitemShut {NoStop}%
\bibitem [{\citenamefont {Aristov}\ \emph {et~al.}(1997)\citenamefont
  {Aristov}, \citenamefont {Maleyev},\ and\ \citenamefont
  {Yashenkin}}]{Aristov97}%
  \BibitemOpen
  \bibfield  {author} {\bibinfo {author} {\bibfnamefont {D.~N.}\ \bibnamefont
  {Aristov}}, \bibinfo {author} {\bibfnamefont {S.~V.}\ \bibnamefont
  {Maleyev}}, \ and\ \bibinfo {author} {\bibfnamefont {A.~G.}\ \bibnamefont
  {Yashenkin}},\ }\href {\doibase 10.1007/s002570050313} {\bibfield  {journal}
  {\bibinfo  {journal} {Zeitschrift f{\"u}r Physik B Condensed Matter}\
  }\textbf {\bibinfo {volume} {102}},\ \bibinfo {pages} {467} (\bibinfo {year}
  {1997})}\BibitemShut {NoStop}%
\bibitem [{\citenamefont {Galitski}\ and\ \citenamefont
  {Larkin}(2002)}]{Galitski02}%
  \BibitemOpen
  \bibfield  {author} {\bibinfo {author} {\bibfnamefont {V.~M.}\ \bibnamefont
  {Galitski}}\ and\ \bibinfo {author} {\bibfnamefont {A.~I.}\ \bibnamefont
  {Larkin}},\ }\href {\doibase 10.1103/PhysRevB.66.064526} {\bibfield
  {journal} {\bibinfo  {journal} {Phys. Rev. B}\ }\textbf {\bibinfo {volume}
  {66}},\ \bibinfo {pages} {064526} (\bibinfo {year} {2002})}\BibitemShut
  {NoStop}%
\bibitem [{\citenamefont {Yao}\ \emph {et~al.}(2014)\citenamefont {Yao},
  \citenamefont {Glazman}, \citenamefont {Demler}, \citenamefont {Lukin},\ and\
  \citenamefont {Sau}}]{Yao14}%
  \BibitemOpen
  \bibfield  {author} {\bibinfo {author} {\bibfnamefont {N.~Y.}\ \bibnamefont
  {Yao}}, \bibinfo {author} {\bibfnamefont {L.~I.}\ \bibnamefont {Glazman}},
  \bibinfo {author} {\bibfnamefont {E.~A.}\ \bibnamefont {Demler}}, \bibinfo
  {author} {\bibfnamefont {M.~D.}\ \bibnamefont {Lukin}}, \ and\ \bibinfo
  {author} {\bibfnamefont {J.~D.}\ \bibnamefont {Sau}},\ }\href {\doibase
  10.1103/PhysRevLett.113.087202} {\bibfield  {journal} {\bibinfo  {journal}
  {Phys. Rev. Lett.}\ }\textbf {\bibinfo {volume} {113}},\ \bibinfo {pages}
  {087202} (\bibinfo {year} {2014})}\BibitemShut {NoStop}%
\bibitem [{\citenamefont {Schecter}\ \emph {et~al.}(2016)\citenamefont
  {Schecter}, \citenamefont {Flensberg}, \citenamefont {Christensen},
  \citenamefont {Andersen},\ and\ \citenamefont {Paaske}}]{Schecter16}%
  \BibitemOpen
  \bibfield  {author} {\bibinfo {author} {\bibfnamefont {M.}~\bibnamefont
  {Schecter}}, \bibinfo {author} {\bibfnamefont {K.}~\bibnamefont {Flensberg}},
  \bibinfo {author} {\bibfnamefont {M.~H.}\ \bibnamefont {Christensen}},
  \bibinfo {author} {\bibfnamefont {B.~M.}\ \bibnamefont {Andersen}}, \ and\
  \bibinfo {author} {\bibfnamefont {J.}~\bibnamefont {Paaske}},\ }\href
  {\doibase 10.1103/PhysRevB.93.140503} {\bibfield  {journal} {\bibinfo
  {journal} {Phys. Rev. B}\ }\textbf {\bibinfo {volume} {93}},\ \bibinfo
  {pages} {140503} (\bibinfo {year} {2016})}\BibitemShut {NoStop}%
\bibitem [{\citenamefont {Christensen}\ \emph {et~al.}(2016)\citenamefont
  {Christensen}, \citenamefont {Schecter}, \citenamefont {Flensberg},
  \citenamefont {Andersen},\ and\ \citenamefont {Paaske}}]{Christensen16}%
  \BibitemOpen
  \bibfield  {author} {\bibinfo {author} {\bibfnamefont {M.~H.}\ \bibnamefont
  {Christensen}}, \bibinfo {author} {\bibfnamefont {M.}~\bibnamefont
  {Schecter}}, \bibinfo {author} {\bibfnamefont {K.}~\bibnamefont {Flensberg}},
  \bibinfo {author} {\bibfnamefont {B.~M.}\ \bibnamefont {Andersen}}, \ and\
  \bibinfo {author} {\bibfnamefont {J.}~\bibnamefont {Paaske}},\ }\href
  {\doibase 10.1103/PhysRevB.94.144509} {\bibfield  {journal} {\bibinfo
  {journal} {Phys. Rev. B}\ }\textbf {\bibinfo {volume} {94}},\ \bibinfo
  {pages} {144509} (\bibinfo {year} {2016})}\BibitemShut {NoStop}%
\bibitem [{\citenamefont {Chandra}\ \emph {et~al.}(1990)\citenamefont
  {Chandra}, \citenamefont {Coleman},\ and\ \citenamefont
  {Larkin}}]{Chandra90}%
  \BibitemOpen
  \bibfield  {author} {\bibinfo {author} {\bibfnamefont {P.}~\bibnamefont
  {Chandra}}, \bibinfo {author} {\bibfnamefont {P.}~\bibnamefont {Coleman}}, \
  and\ \bibinfo {author} {\bibfnamefont {A.~I.}\ \bibnamefont {Larkin}},\
  }\href {\doibase 10.1103/PhysRevLett.64.88} {\bibfield  {journal} {\bibinfo
  {journal} {Phys. Rev. Lett.}\ }\textbf {\bibinfo {volume} {64}},\ \bibinfo
  {pages} {88} (\bibinfo {year} {1990})}\BibitemShut {NoStop}%
\bibitem [{\citenamefont {Capriotti}\ and\ \citenamefont
  {Sachdev}(2004)}]{Capriotti04}%
  \BibitemOpen
  \bibfield  {author} {\bibinfo {author} {\bibfnamefont {L.}~\bibnamefont
  {Capriotti}}\ and\ \bibinfo {author} {\bibfnamefont {S.}~\bibnamefont
  {Sachdev}},\ }\href {\doibase 10.1103/PhysRevLett.93.257206} {\bibfield
  {journal} {\bibinfo  {journal} {Phys. Rev. Lett.}\ }\textbf {\bibinfo
  {volume} {93}},\ \bibinfo {pages} {257206} (\bibinfo {year}
  {2004})}\BibitemShut {NoStop}%
\bibitem [{\citenamefont {Weber}\ \emph {et~al.}(2003)\citenamefont {Weber},
  \citenamefont {Capriotti}, \citenamefont {Misguich}, \citenamefont {Becca},
  \citenamefont {Elhajal},\ and\ \citenamefont {Mila}}]{Weber03}%
  \BibitemOpen
  \bibfield  {author} {\bibinfo {author} {\bibfnamefont {C.}~\bibnamefont
  {Weber}}, \bibinfo {author} {\bibfnamefont {L.}~\bibnamefont {Capriotti}},
  \bibinfo {author} {\bibfnamefont {G.}~\bibnamefont {Misguich}}, \bibinfo
  {author} {\bibfnamefont {F.}~\bibnamefont {Becca}}, \bibinfo {author}
  {\bibfnamefont {M.}~\bibnamefont {Elhajal}}, \ and\ \bibinfo {author}
  {\bibfnamefont {F.}~\bibnamefont {Mila}},\ }\href {\doibase
  10.1103/PhysRevLett.91.177202} {\bibfield  {journal} {\bibinfo  {journal}
  {Phys. Rev. Lett.}\ }\textbf {\bibinfo {volume} {91}},\ \bibinfo {pages}
  {177202} (\bibinfo {year} {2003})}\BibitemShut {NoStop}%
\bibitem [{\citenamefont {{Villain, J.}}(1977)}]{Villain77}%
  \BibitemOpen
  \bibfield  {author} {\bibinfo {author} {\bibnamefont {{Villain, J.}}},\
  }\href {\doibase 10.1051/jphys:01977003804038500} {\bibfield  {journal}
  {\bibinfo  {journal} {J. Phys. France}\ }\textbf {\bibinfo {volume} {38}},\
  \bibinfo {pages} {385} (\bibinfo {year} {1977})}\BibitemShut {NoStop}%
\bibitem [{\citenamefont {Henley}(1989)}]{Henley89}%
  \BibitemOpen
  \bibfield  {author} {\bibinfo {author} {\bibfnamefont {C.~L.}\ \bibnamefont
  {Henley}},\ }\href {\doibase 10.1103/PhysRevLett.62.2056} {\bibfield
  {journal} {\bibinfo  {journal} {Phys. Rev. Lett.}\ }\textbf {\bibinfo
  {volume} {62}},\ \bibinfo {pages} {2056} (\bibinfo {year}
  {1989})}\BibitemShut {NoStop}%
\bibitem [{\citenamefont {Schecter}\ \emph {et~al.}(2017)\citenamefont
  {Schecter}, \citenamefont {Sylju\aa{}sen},\ and\ \citenamefont
  {Paaske}}]{Schecter17}%
  \BibitemOpen
  \bibfield  {author} {\bibinfo {author} {\bibfnamefont {M.}~\bibnamefont
  {Schecter}}, \bibinfo {author} {\bibfnamefont {O.~F.}\ \bibnamefont
  {Sylju\aa{}sen}}, \ and\ \bibinfo {author} {\bibfnamefont {J.}~\bibnamefont
  {Paaske}},\ }\href {\doibase 10.1103/PhysRevLett.119.157202} {\bibfield
  {journal} {\bibinfo  {journal} {Phys. Rev. Lett.}\ }\textbf {\bibinfo
  {volume} {119}},\ \bibinfo {pages} {157202} (\bibinfo {year}
  {2017})}\BibitemShut {NoStop}%
\bibitem [{\citenamefont {Chakravarty}\ \emph {et~al.}(1989)\citenamefont
  {Chakravarty}, \citenamefont {Halperin},\ and\ \citenamefont
  {Nelson}}]{Chakravarty89}%
  \BibitemOpen
  \bibfield  {author} {\bibinfo {author} {\bibfnamefont {S.}~\bibnamefont
  {Chakravarty}}, \bibinfo {author} {\bibfnamefont {B.~I.}\ \bibnamefont
  {Halperin}}, \ and\ \bibinfo {author} {\bibfnamefont {D.~R.}\ \bibnamefont
  {Nelson}},\ }\href {\doibase 10.1103/PhysRevB.39.2344} {\bibfield  {journal}
  {\bibinfo  {journal} {Phys. Rev. B}\ }\textbf {\bibinfo {volume} {39}},\
  \bibinfo {pages} {2344} (\bibinfo {year} {1989})}\BibitemShut {NoStop}%
\bibitem [{\citenamefont {Capriotti}\ \emph {et~al.}(2004)\citenamefont
  {Capriotti}, \citenamefont {Fubini}, \citenamefont {Roscilde},\ and\
  \citenamefont {Tognetti}}]{Capriotti04b}%
  \BibitemOpen
  \bibfield  {author} {\bibinfo {author} {\bibfnamefont {L.}~\bibnamefont
  {Capriotti}}, \bibinfo {author} {\bibfnamefont {A.}~\bibnamefont {Fubini}},
  \bibinfo {author} {\bibfnamefont {T.}~\bibnamefont {Roscilde}}, \ and\
  \bibinfo {author} {\bibfnamefont {V.}~\bibnamefont {Tognetti}},\ }\href
  {\doibase 10.1103/PhysRevLett.92.157202} {\bibfield  {journal} {\bibinfo
  {journal} {Phys. Rev. Lett.}\ }\textbf {\bibinfo {volume} {92}},\ \bibinfo
  {pages} {157202} (\bibinfo {year} {2004})}\BibitemShut {NoStop}%
\bibitem [{\citenamefont {Luh}(1965)}]{Luh65}%
  \BibitemOpen
  \bibfield  {author} {\bibinfo {author} {\bibfnamefont {Y.}~\bibnamefont
  {Luh}},\ }\href@noop {} {\bibfield  {journal} {\bibinfo  {journal} {Acta
  Phys. Sin.}\ }\textbf {\bibinfo {volume} {21}},\ \bibinfo {pages} {75}
  (\bibinfo {year} {1965})}\BibitemShut {NoStop}%
\bibitem [{\citenamefont {Shiba}(1968)}]{Shiba68}%
  \BibitemOpen
  \bibfield  {author} {\bibinfo {author} {\bibfnamefont {H.}~\bibnamefont
  {Shiba}},\ }\href {\doibase 10.1143/PTP.40.435} {\bibfield  {journal}
  {\bibinfo  {journal} {Progress of Theoretical Physics}\ }\textbf {\bibinfo
  {volume} {40}},\ \bibinfo {pages} {435} (\bibinfo {year} {1968})}\BibitemShut
  {NoStop}%
\bibitem [{\citenamefont {Rusinov}(1969)}]{Rusinov69}%
  \BibitemOpen
  \bibfield  {author} {\bibinfo {author} {\bibfnamefont {A.~I.}\ \bibnamefont
  {Rusinov}},\ }\href@noop {} {\bibfield  {journal} {\bibinfo  {journal} {JETP
  Lett.}\ }\textbf {\bibinfo {volume} {9}},\ \bibinfo {pages} {85} (\bibinfo
  {year} {1969})}\BibitemShut {NoStop}%
\bibitem [{\citenamefont {Anderson}\ and\ \citenamefont
  {Suhl}(1959)}]{Anderson59}%
  \BibitemOpen
  \bibfield  {author} {\bibinfo {author} {\bibfnamefont {P.~W.}\ \bibnamefont
  {Anderson}}\ and\ \bibinfo {author} {\bibfnamefont {H.}~\bibnamefont
  {Suhl}},\ }\href {\doibase 10.1103/PhysRev.116.898} {\bibfield  {journal}
  {\bibinfo  {journal} {Phys. Rev.}\ }\textbf {\bibinfo {volume} {116}},\
  \bibinfo {pages} {898} (\bibinfo {year} {1959})}\BibitemShut {NoStop}%
\bibitem [{Note1()}]{Note1}%
  \BibitemOpen
  \bibinfo {note} {See Supplemental Material at [URL will be inserted by
  publisher] for some typical spin configurations in these symmetry broken
  regions.}\BibitemShut {Stop}%
\bibitem [{\citenamefont {Seabra}\ \emph {et~al.}(2016)\citenamefont {Seabra},
  \citenamefont {Sindzingre}, \citenamefont {Momoi},\ and\ \citenamefont
  {Shannon}}]{Seabra16}%
  \BibitemOpen
  \bibfield  {author} {\bibinfo {author} {\bibfnamefont {L.}~\bibnamefont
  {Seabra}}, \bibinfo {author} {\bibfnamefont {P.}~\bibnamefont {Sindzingre}},
  \bibinfo {author} {\bibfnamefont {T.}~\bibnamefont {Momoi}}, \ and\ \bibinfo
  {author} {\bibfnamefont {N.}~\bibnamefont {Shannon}},\ }\href {\doibase
  10.1103/PhysRevB.93.085132} {\bibfield  {journal} {\bibinfo  {journal} {Phys.
  Rev. B}\ }\textbf {\bibinfo {volume} {93}},\ \bibinfo {pages} {085132}
  (\bibinfo {year} {2016})}\BibitemShut {NoStop}%
\end{thebibliography}%

\pagebreak
\onecolumngrid
\begin{center}
\textbf{\large Supplemental Material}
\\
\bigskip
\textbf{\large Cooper Pair Induced Frustration and Nematicity of Two-Dimensional Magnetic Adatom Lattices}
\end{center}
\setcounter{equation}{0}
\setcounter{figure}{0}
\setcounter{table}{0}
\setcounter{page}{1}
\makeatletter

\renewcommand{\theequation}{S\arabic{equation}}
\renewcommand{\thefigure}{S\arabic{figure}}
\renewcommand{\bibnumfmt}[1]{[S#1]}
\renewcommand{\citenumfont}[1]{S#1}

In this supplemental materials section we present some typical low free-energy spin configurations for the effective adatom spin Hamiltonian, Eqs.~(2) and (3) in the main paper. The spin configurations are obtained using Monte Carlo simulations on a square lattice of size $L \times L$ with open boundary conditions. In all the results presented below we use $\epsilon/\Delta=0.5$ and $\Delta/E_F = 5 \times 10^{-3}$, the same as in the Figs.~2-4 in the main paper.  

\begin{figure}[htb]
\begin{center}
\mbox{
    \includegraphics[clip,width=5cm]{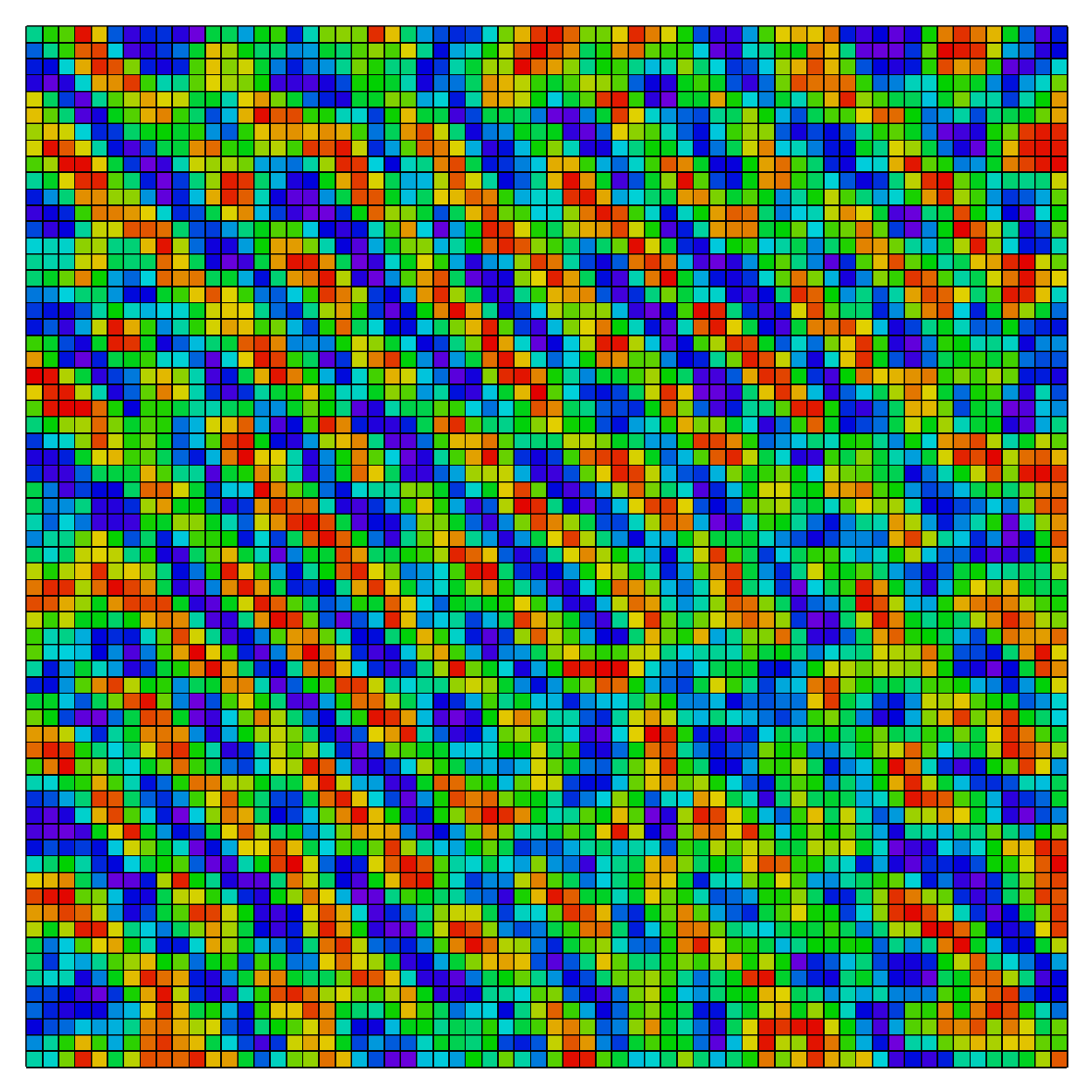}  
   \includegraphics[clip,width=5cm]{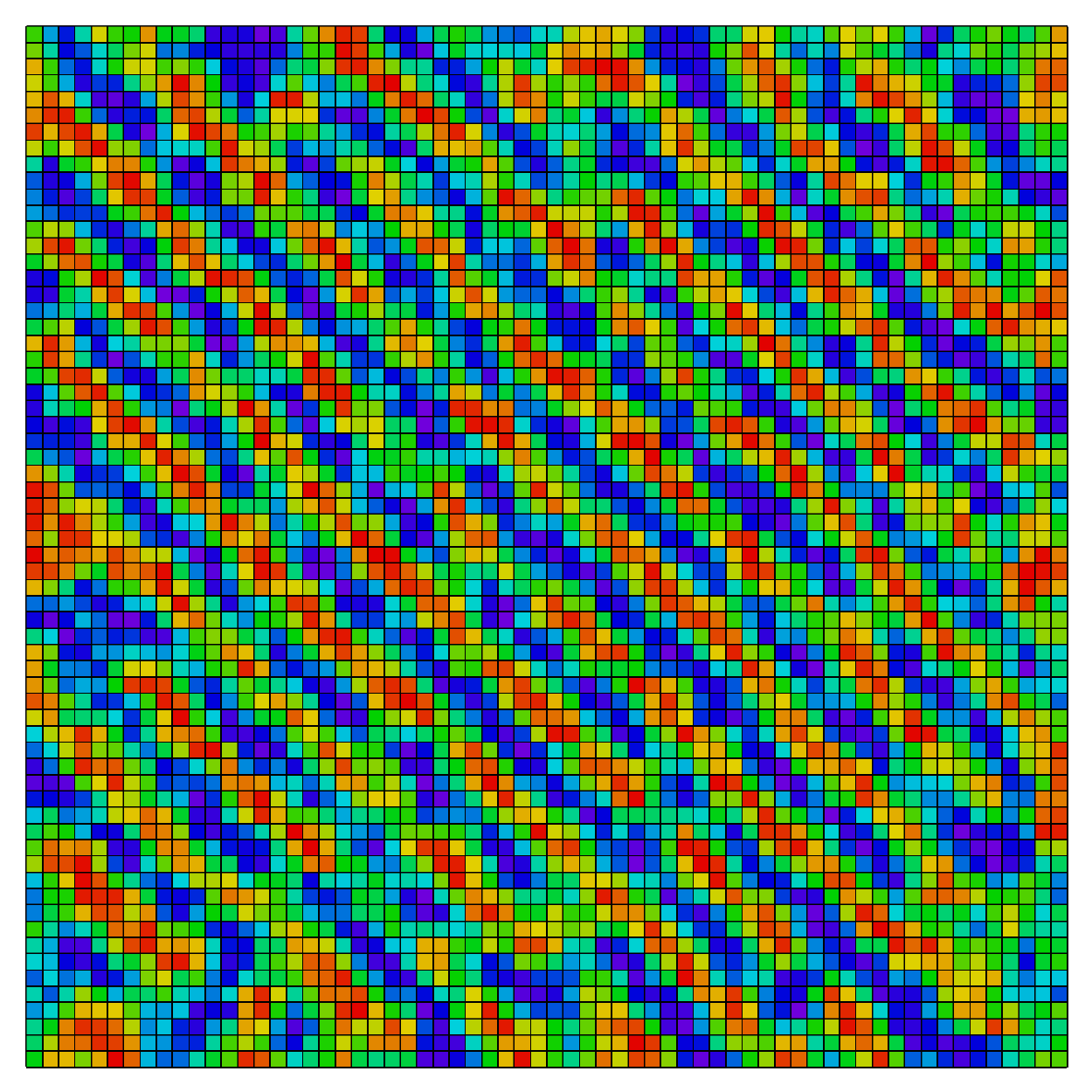}  
   \includegraphics[clip,width=5cm]{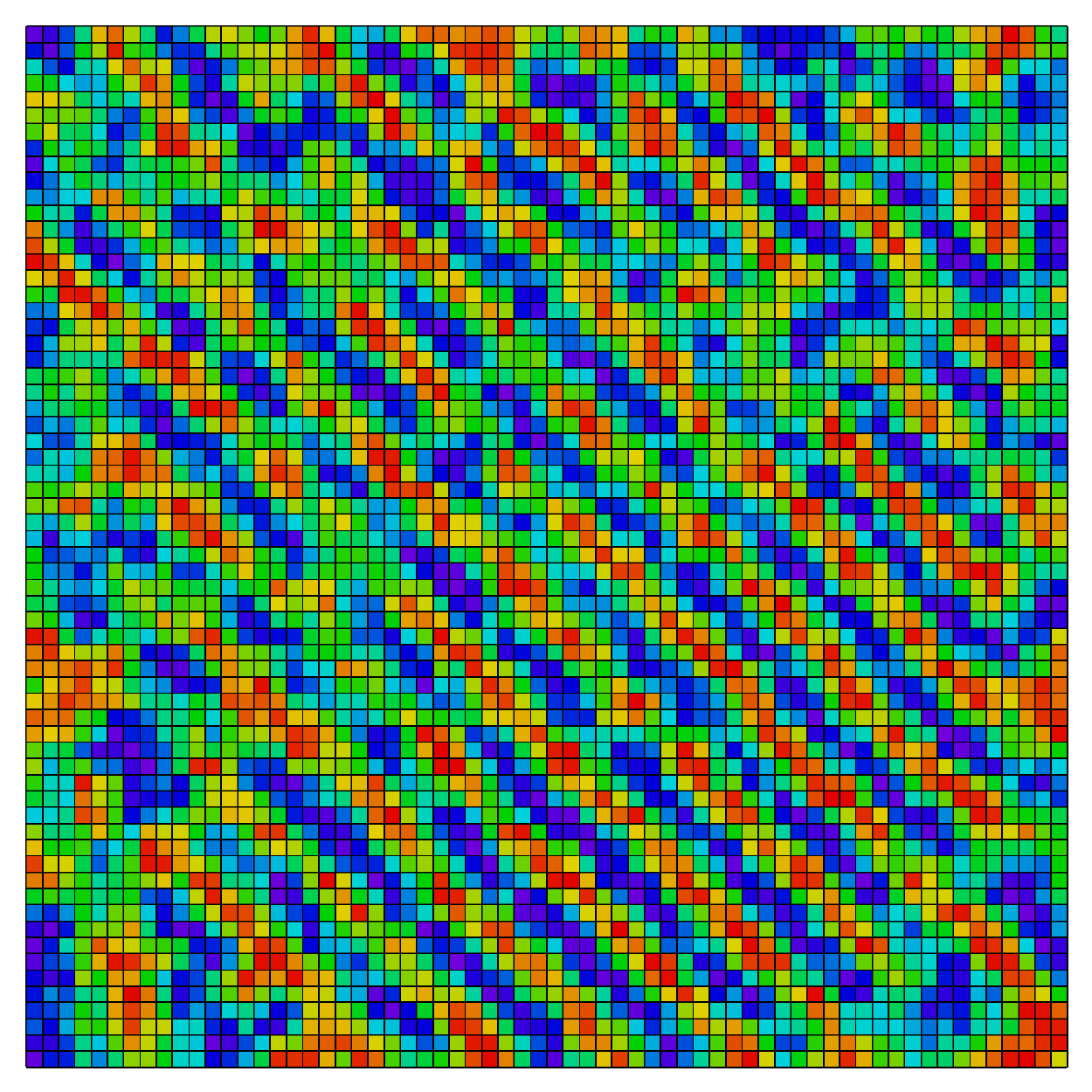}}  
\includegraphics[clip,width=5cm]{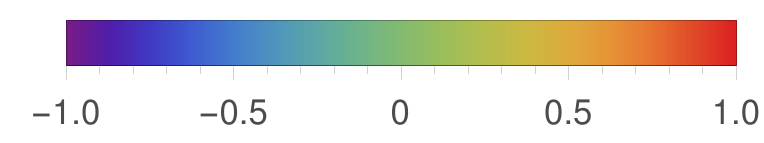}  
\caption{(color online) Snapshot of a single spin configuration for $k_F a/\pi = 3.44$, $T/\Delta=1.6 \times 10^{-3}$, $L=64$. Each panel shows one spin-component: $S_x$(left) , $S_y$ (middle), and $S_z$(right).} \label{spins344}
\end{center}
\end{figure}

Fig.~\ref{spins344} shows a single typical spin configuration obtained from the Monte Carlo simulation at $k_F a/\pi = 3.44$ at a low temperature. Each panel shows the spatial distribution of a single spin-component which takes values in the interval $[-1,1]$ indicated by the colorscale. One can clearly see diagonal structures with a $\vec{Q}=(+q,+q)$ consistent with the blue $\sigma^d \neq 0$ region in the phase diagram in Fig.~2 in the main paper. Although not so clearly visible, it also appears for this configuration that the spins lie predominantly in the yz-plane below the main diagonal of the lattice and changes into the xy-plane above the diagonal.  

\begin{figure}[htb]
\begin{center}
\mbox{
    \includegraphics[clip,width=5cm]{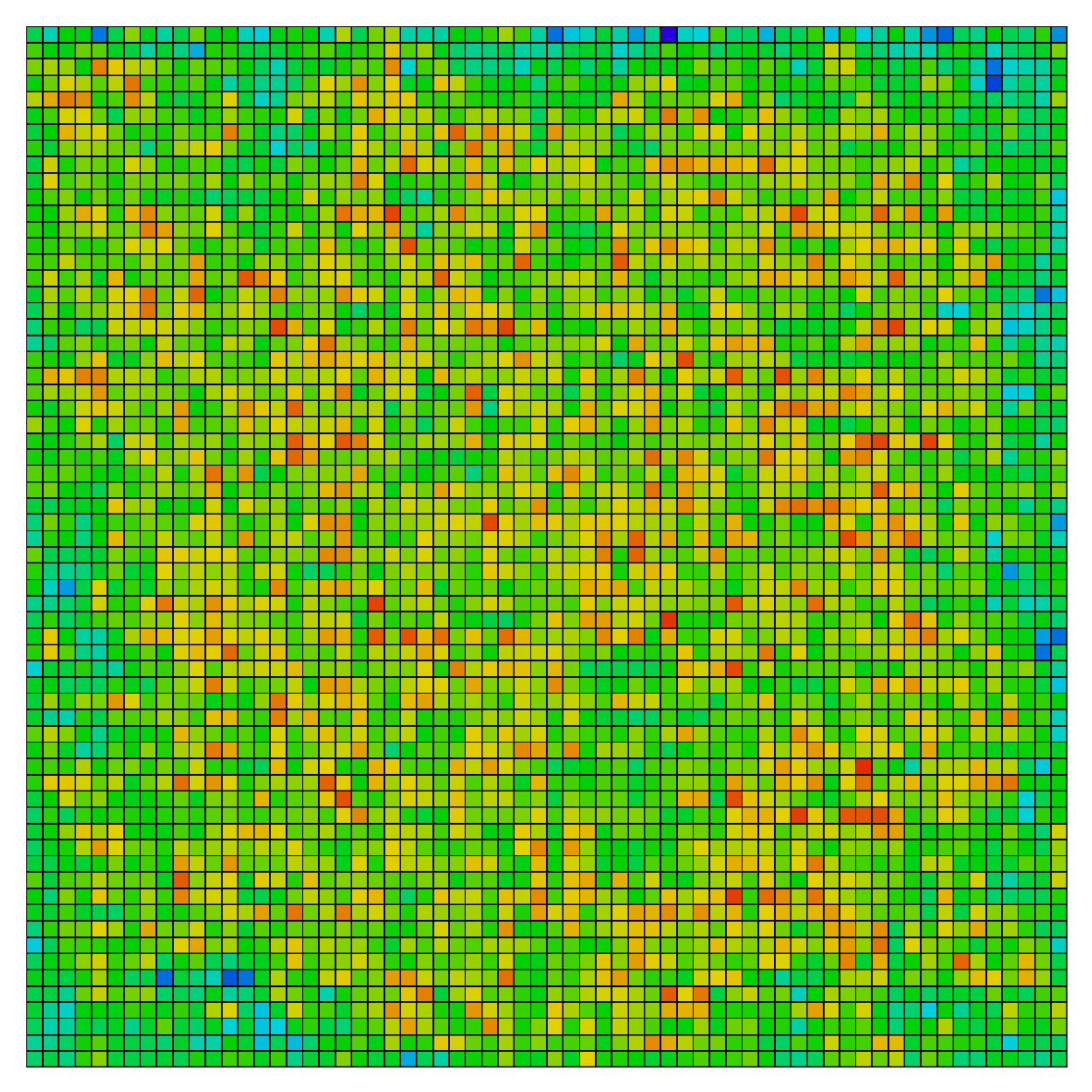}  
   \includegraphics[clip,width=5cm]{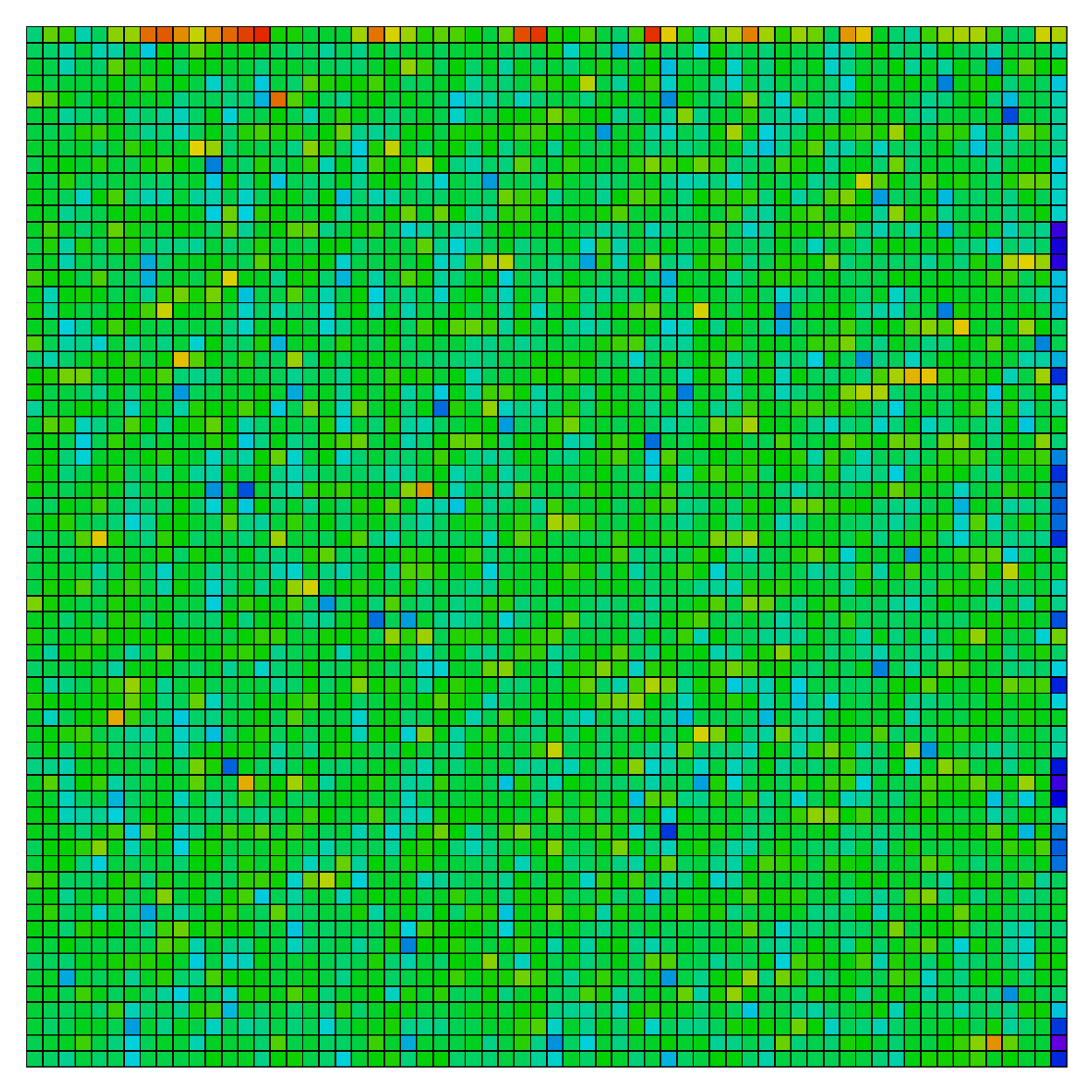}}

\includegraphics[clip,width=5cm]{colorbar.pdf}
\caption{(color online) Spatial distribution for the two plaquette order parameters. $\sigma^d_{\vec{r}}$ (left) and $\sigma^a_{\vec{r}}$ (right) for the same spin configuration as in Fig.~\ref{spins344}. \label{plaq344}}
\end{center}
\end{figure}

In Fig.~\ref{plaq344} we have also plotted the spatial distributions of the local plaquette order parameters $\sigma^d_{\vec{r}}$ and $\sigma^a_{\vec{r}}$
\begin{align}
     \sigma^{d}_{\vec{r}} &= \frac{1}{2} \left( 
\vec{S}_{\vec{r}+\hat{y}} \cdot \vec{S}_{\vec{r}+\hat{x}} 
-\vec{S}_{\vec{r}} \cdot \vec{S}_{\vec{r}+\hat{x}+\hat{y}} 
\right) \label{sigmad} \\
     \sigma^{a}_{\vec{r}} &= \frac{1}{2} \left( \vec{S}_{\vec{r}} \cdot \vec{S}_{\vec{r}+\hat{x}} - \vec{S}_{\vec{r}} \cdot \vec{S}_{\vec{r}+\hat{y}} 
\right) \label{sigmaa}
\end{align}
 for the same spin configuration as in Fig.~\ref{spins344}. These local plaquette order parameters Eqs.~(\ref{sigmad}),(\ref{sigmaa}) transform in the same way under lattice transformations as the order parameters in Eq.~(5) in the main paper. 
The diagonal order parameter $\sigma^d$ develops a positive magnitude over a large area consistent with a $\vec{Q}$ of the form $(+q,+q)$. At the very top (right) boundary it is also seen that $\sigma^a$ becomes non-zero which is a consequence of the open boundary condition.

\begin{figure}[htb]
\begin{center}
\mbox{
    \includegraphics[clip,width=5cm]{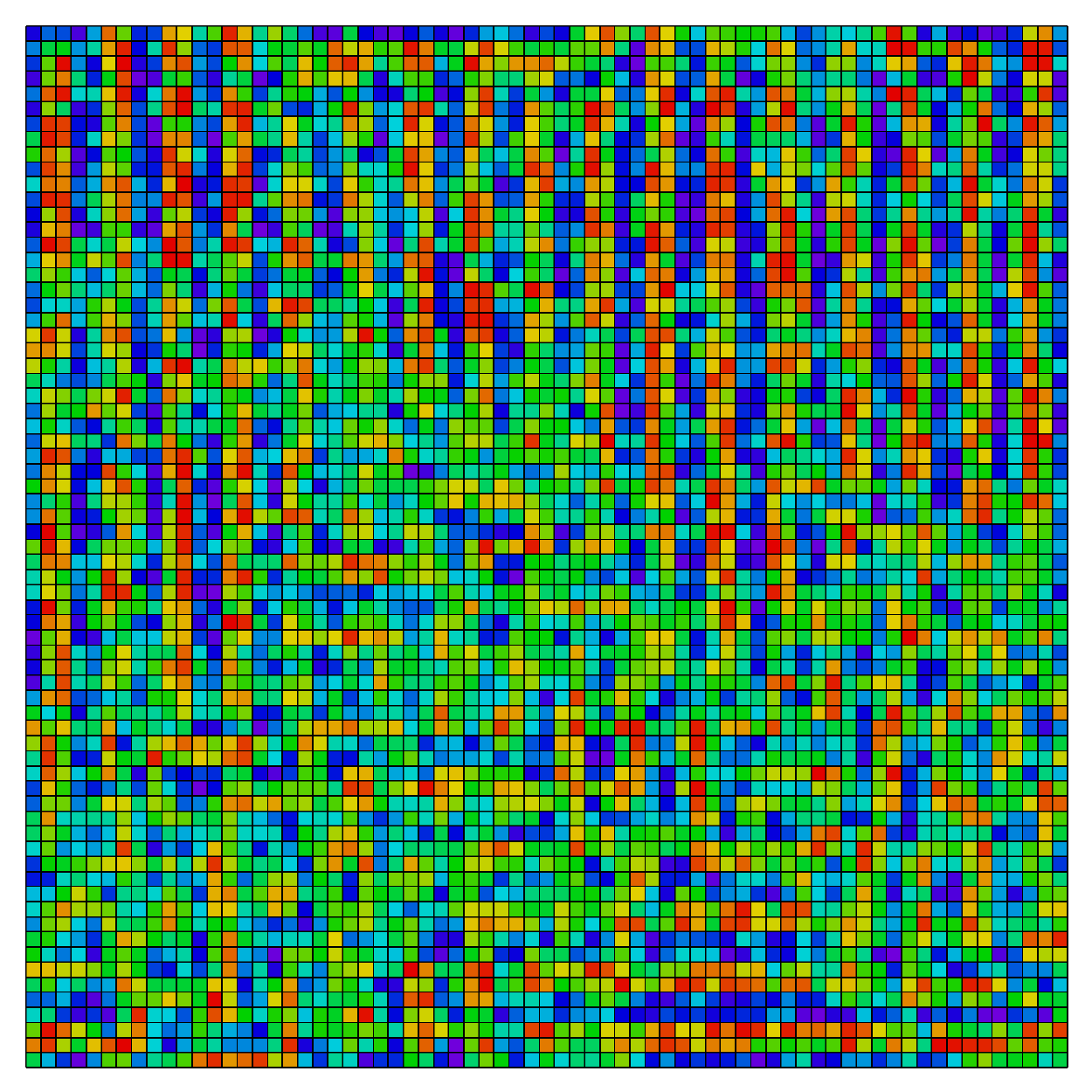}  
   \includegraphics[clip,width=5cm]{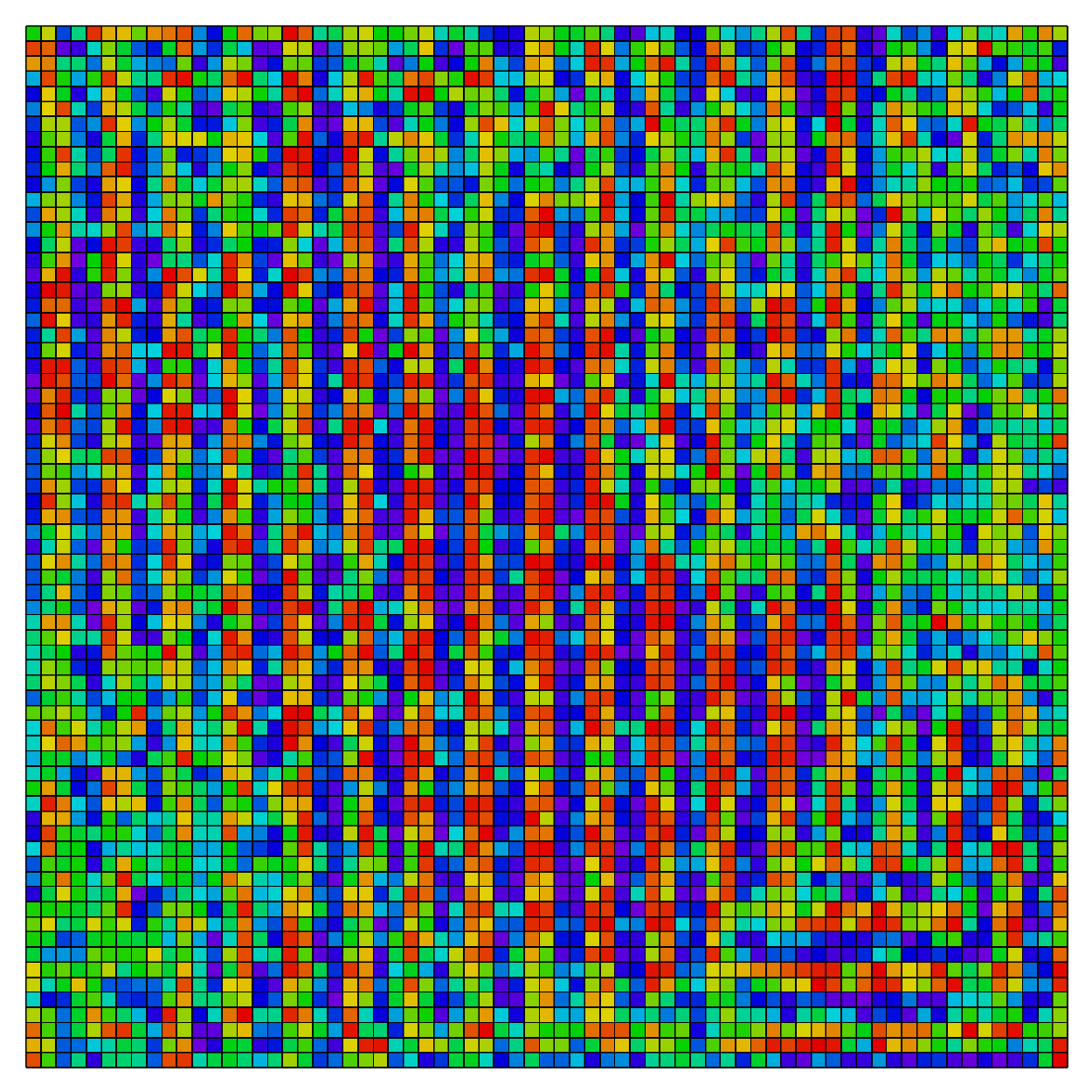}  
   \includegraphics[clip,width=5cm]{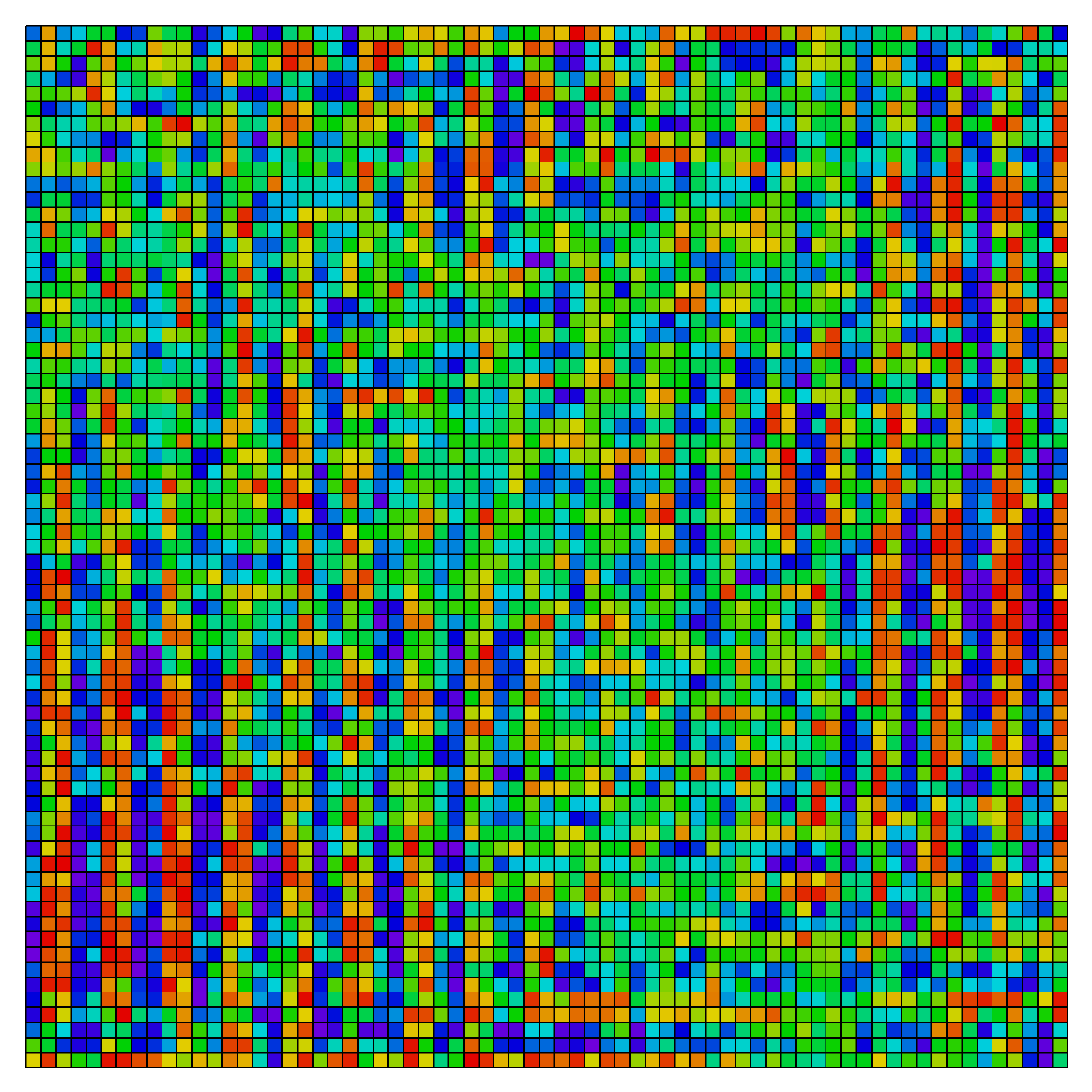}}  
\includegraphics[clip,width=5cm]{colorbar.pdf}  
\caption{(color online) Snapshot of a single spin configuration for $k_F a/\pi = 3.6$, $T/\Delta=2.5 \times 10^{-3}$, $L=69$.  Each panel shows one spin-component: $S_x$(left) , $S_y$ (middle), and $S_z$(right).} \label{spins36}
\end{center}
\end{figure}

\begin{figure}[htb]
\begin{center}
\mbox{
    \includegraphics[clip,width=5cm]{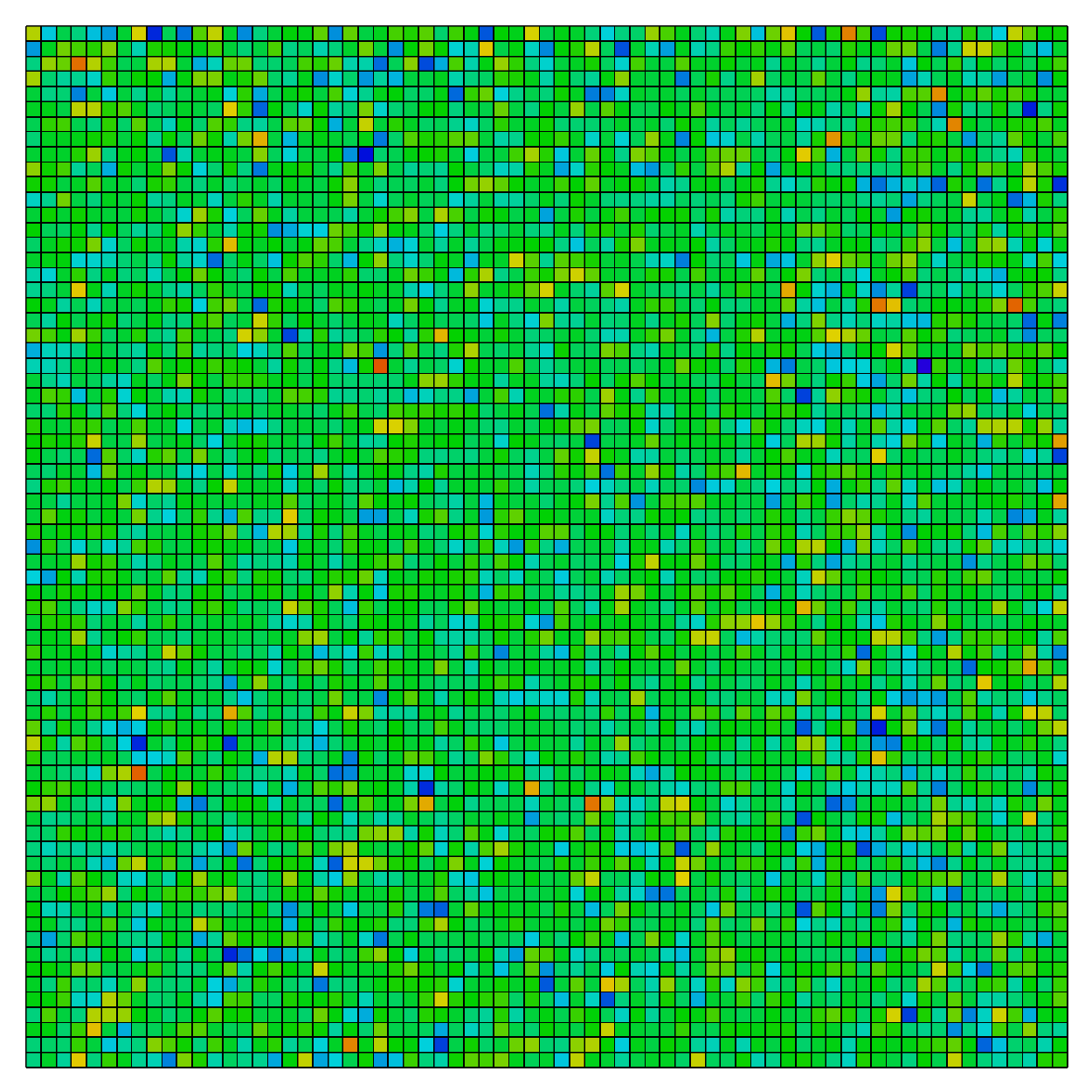}  
   \includegraphics[clip,width=5cm]{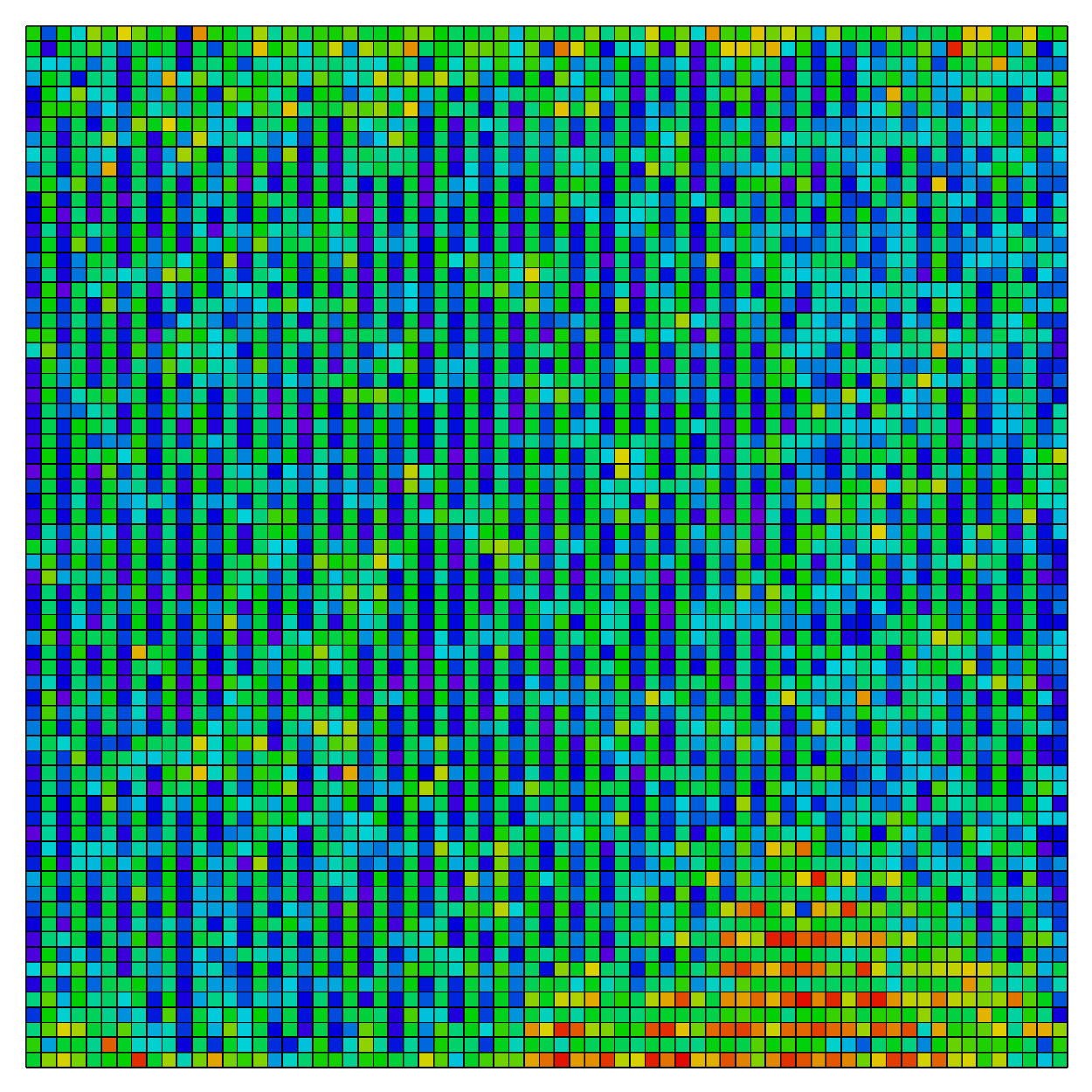}}

\includegraphics[clip,width=5cm]{colorbar.pdf}
\caption{(color online) Spatial distribution for the two local plaquette order parameters. $\sigma^d_{\vec{r}}$ (left) and $\sigma^a_{\vec{r}}$ (right) for the same spin configuration as in Fig.~\ref{spins36}. \label{plaq36}}
\end{center}
\end{figure}

For a slightly larger value of $k_F a$ at low $T$ one reaches a phase where the preferred $\vec{Q}$ lies along the axes(red region in Fig.~2 in the main paper). Fig.~\ref{spins36} shows the spin components for a single typical low temperature spin configuration at $k_F a/\pi=3.6$. In most of the lattice, except at the bottom on the right, there are vertical stripes with a $\vec{Q}$-vector along the x-axis. One can clearly see the rapid modulation along the x-axis consistent with Fig.~3 in the main paper. At the bottom on the right side there is a small region with horizontal stripes. This region becomes more visible in the $\sigma^a$-order parameter plot for this configuration shown in Fig.~\ref{plaq36}. There the $\sigma^a$ order parameter is negative over most of the lattice consistent with vertical stripes, while it is positive in a smaller region at the bottom right. Such minority phase regions disappear as the temperature is lowered further(not shown). In contrast to Fig.~\ref{plaq344} where there is no apparent spatial structure to the local plaquette order parameter $\sigma^d$, the $\sigma^a_{\vec{r}}$ order parameter in Fig.~\ref{plaq36} displays a clear spatial modulation indicative of the breaking of both rotational and translational symmetry.

\clearpage

\subsection*{Description of the Monte Carlo method}
The Monte Carlo algorithm employed here uses both standard Metropolis moves as well as overrelaxation moves.
In a Monte Carlo move of either type a site is first selected at random. Then the effective site magnetic field from all other spins at the selected site is computed. The long-range nature of the interaction makes this computation costly in terms of computing time as all spins on the lattice must be visited. In the overrelaxation move, the spin is then rotated a random angle about the direction of this effective site magnetic field. As this rotation does not change the energy, it is always accepted. In the Metropolis accept/reject move, the spin is reflected about the plane which normal vector is the effective site magnetic field. A Monte Carlo sweep(MCS) contains $N=L^2$ Metropolis moves followed by another $N$ overrelaxation moves. Typically $10^5$ MCS are used to equilibrate the system before the spin configuration is recorded.

We have used open boundary conditions for two reasons. First it is the most relevant boundary condition for the experimental situation. Second it does not require any fine-tuning of the lattice size $L$. For periodic boundary conditions the system size $L$ must be chosen carefully to accomodate spirals with the lowest energy wavevector $\vec{Q}$ so as to avoid strains due to the boundary condition. This choice of lattice size is also made more difficult by the long-range form of the interaction which causes the value of $\vec{Q}$ to depend quite significantly on $L$ for the parameters and system sizes chosen here.

\end{document}